\documentclass[
    twocolumn,
	prl,
	amssymb,
	preprintnumbers,
	secnumarabic,
	nofootinbib,
	superscriptaddress,
	]{revtex4-1}
	
\pdfoutput=1

\usepackage{hyperref}
\usepackage{feynmf}
\usepackage{graphicx}
\usepackage{enumitem}
\usepackage{latexsym}
\usepackage{amsfonts}
\usepackage{amssymb}
\usepackage{color}
\usepackage{amsmath}
\usepackage{slashed}
\usepackage{dcolumn}
\usepackage{verbatim}
\usepackage{comment}
\usepackage{float}
\usepackage{multirow}
\usepackage{xspace}
\usepackage[normalem]{ulem}
\graphicspath{ {./Figs}{./} }

\newcommand{\beq}{\begin{eqnarray}}
\newcommand{\eeq}{\end{eqnarray}}
\newcommand{\bea}{\begin{eqnarray}}
\newcommand{\eea}{\end{eqnarray}}

\newcommand{\kev}{\text{keV}}
\newcommand{\mev}{\text{MeV}}
\newcommand{\gev}{\text{GeV}}
\newcommand{\tev}{\text{TeV}}
\newcommand{\lrf}[2]{\!\left(\frac{#1}{#2}\right)\!}

\allowdisplaybreaks

\begin{document}

\title{Dark Matter Direct Detection on the Moon}

\author{Andr\'ea Gaspert}
\email{agaspert@stanford.edu}
\affiliation{Department of Physics, Stanford University, Stanford, CA 94305, USA}

\author{Pietro Giampa}
\email{pgiampa@triumf.ca}
\affiliation{TRIUMF, 4004 Wesbrook Mall, Vancouver, BC V6T 2A3, Canada}

\author{Navin McGinnis}
\email{nmcginnis@triumf.ca}
\affiliation{TRIUMF, 4004 Wesbrook Mall, Vancouver, BC V6T 2A3, Canada}

\author{David E. Morrissey}
\email{dmorri@triumf.ca}
\affiliation{TRIUMF, 4004 Wesbrook Mall, Vancouver, BC V6T 2A3, Canada}

\begin{abstract}
Direct searches for dark matter with large-scale noble liquid detectors have become sensitive enough to detect the coherent scattering of local neutrinos. This will become a very challenging background to dark matter discovery in planned future detectors. For dark matter with mass above $10\,\gev$, the dominant neutrino backgrounds on the Earth are atmospheric neutrinos created by cosmic ray collisions with the atmosphere. In contrast, the Moon has almost no atmosphere and nearly all cosmic rays incident on the Moon first collide with the lunar surface, producing a very different neutrino spectrum. In this work we estimate the total flux and spectrum of neutrinos near the surface of the Moon. We then use this to show that a large-scale liquid xenon or argon detector located on the Moon could potentially have significantly greater sensitivity to dark matter compared to an equivalent detector on the Earth due to effectively reduced neutrino backgrounds.
\end{abstract}

\maketitle

%%%%%%
%\section{Introduction\label{sec:intro}}
{\flushleft{\bf Introduction:}}
%%%%%%
Our nearest celestial neighbor, the Moon, has fascinated humankind since time immemorial~\cite{moon2015,lunarsourcebook}. The Moon has also helped us understand our Universe, such as through tests of gravity with laser lunar ranging~\cite{Dickey:1994zz,Williams:2004qba,Will:2014kxa}. Recent renewed interest in visiting and exploring the Moon~\cite{artemis} has motivated proposals for new scientific facilities to be built there, including instruments for astronomy and cosmology~\cite{Silk:2020bsr,artemissci} and energy-frontier particle colliders~\cite{Beacham:2021lgt}. In this letter we demonstrate that the Moon could also enable the direct detection of dark matter beyond what is possible on the Earth.

Evidence for dark matter~(DM) from astrophysics and cosmology is overwhelming~\cite{Planck:2018vyg}, but its specific identity remains a mystery~\cite{Battaglieri:2017aum}. A leading paradigm for DM is a new elementary particle species $\chi$ that interacts with ordinary matter more strongly than through gravity alone~\cite{Jungman:1995df,Bertone:2004pz,Feng:2010gw,Roszkowski:2017nbc,Lin:2019uvt,Cooley:2022ufh}. For masses in the $1\,\gev$--$50\,\tev$ range, near the electroweak scale, such a particle could be created and obtain the observed DM density by thermal processes in the hot early Universe~\cite{Lee:1977ua}. This paradigm has motivated a worldwide program to search for DM in the lab (on Earth) by its scattering with ordinary matter in deep underground detectors~\cite{Goodman:1984dc,Drukier:1986tm}.

Efforts to identify DM through this \emph{direct detection} method have made enormous progress over the past decades~\cite{Billard:2021uyg}. For DM with mass $m_\chi \gtrsim 10\,\gev$ and primarily spin-independent interactions with nuclei, the most sensitive current experiments are large-volume noble-liquid detectors using xenon~\cite{PandaX-4T:2021bab,LZ:2022ufs,XENON:2023sxq} or argon~\cite{DEAP:2019yzn,Aalseth:2017fik} as the target material. These detectors are so sensitive that they are beginning to observe the coherent scattering of neutrinos on nuclei~\cite{Freedman:1973yd}. While this is a remarkable achievement, it also implies that neutrino scattering will be a difficult background in future direct searches for DM~\cite{Monroe:2007xp,Dodelson:2008yx,Strigari:2009bq,Billard:2013qya,Ruppin:2014bra,OHare:2016pjy,Gelmini:2018ogy,OHare:2020lva,Newstead:2020fie,Gaspert:2021gyj,OHare:2021utq,Akerib:2022ort}. Indeed, proposed detectors such as DARWIN~\cite{DARWIN:2016hyl,Aalbers:2022dzr} and ARGO~\cite{Galbiati:2018} are expected to be able to look for DM all the way down to the \emph{neutrino floor}~\cite{Billard:2013qya,Ruppin:2014bra} (or \emph{neutrino fog}~\cite{OHare:2021utq,Akerib:2022ort}) beyond which neutrino backgrounds make further progress very challenging. Many well-motivated theories of DM predict scattering cross sections below the neutrino floor~\cite{Ellis:2005mb,Cheung:2012qy,Hill:2013hoa,Cahill-Rowley:2014boa,Bramante:2015una,Claude:2021sye}, such as the infamously challenging $m_\chi = 1.1\,\tev$ Higgsino~\cite{Hill:2013hoa,Hill:2014yxa,Co:2021ion,Dessert:2022evk}. Going beyond it would appear to require directional sensitivity~\cite{OHare:2015utx,Mayet:2016zxu,OHare:2022jnx}, combining data from detectors consisting of different target materials~\cite{Ruppin:2014bra,OHare:2020lva,Gaspert:2021gyj}, or extremely large detector volumes~\cite{Ruppin:2014bra,OHare:2020lva,OHare:2021utq}. Probing below the neutrino floor is complicated further by uncertainties in the spectral shapes of neutrino fluxes and the energy dependences of detector responses~\cite{Gaspert:2021gyj}.

In this letter we show that locating a large-scale detector under the surface of the Moon could allow for greater sensitivity to DM by reducing neutrino backgrounds. The dominant neutrino backgrounds for the detection of DM with mass $m_\chi \gtrsim 10\,\gev$ on the Earth are atmospheric neutrinos, created when cosmic ray~(CR) protons and helium collide with molecules in the atmosphere to produce pions and kaons which yield neutrinos in their subsequent decay chains~\cite{Barr:1989ru}. In contrast, the Moon has almost no atmosphere and CRs collide primarily with the thin regolith layer covering the lunar surface or the underlying rocky crust. This greatly alters the resulting neutrino flux spectrum: on the Earth the pions and kaons decay while in flight, whereas on the Moon they are mostly stopped or absorbed before decaying~\cite{Volkova:1965,Miller:2006,Demidov:2020bff}. We find that the modified neutrino spectrum on the Moon makes for a significantly weaker background to DM scattering relative to the Earth and could provide a novel approach to exploring DM below the (Earth) neutrino floor.

\begin{figure}[t!]
    \centering
%\makebox[\textwidth][c]{  
\includegraphics[width=.47\textwidth]{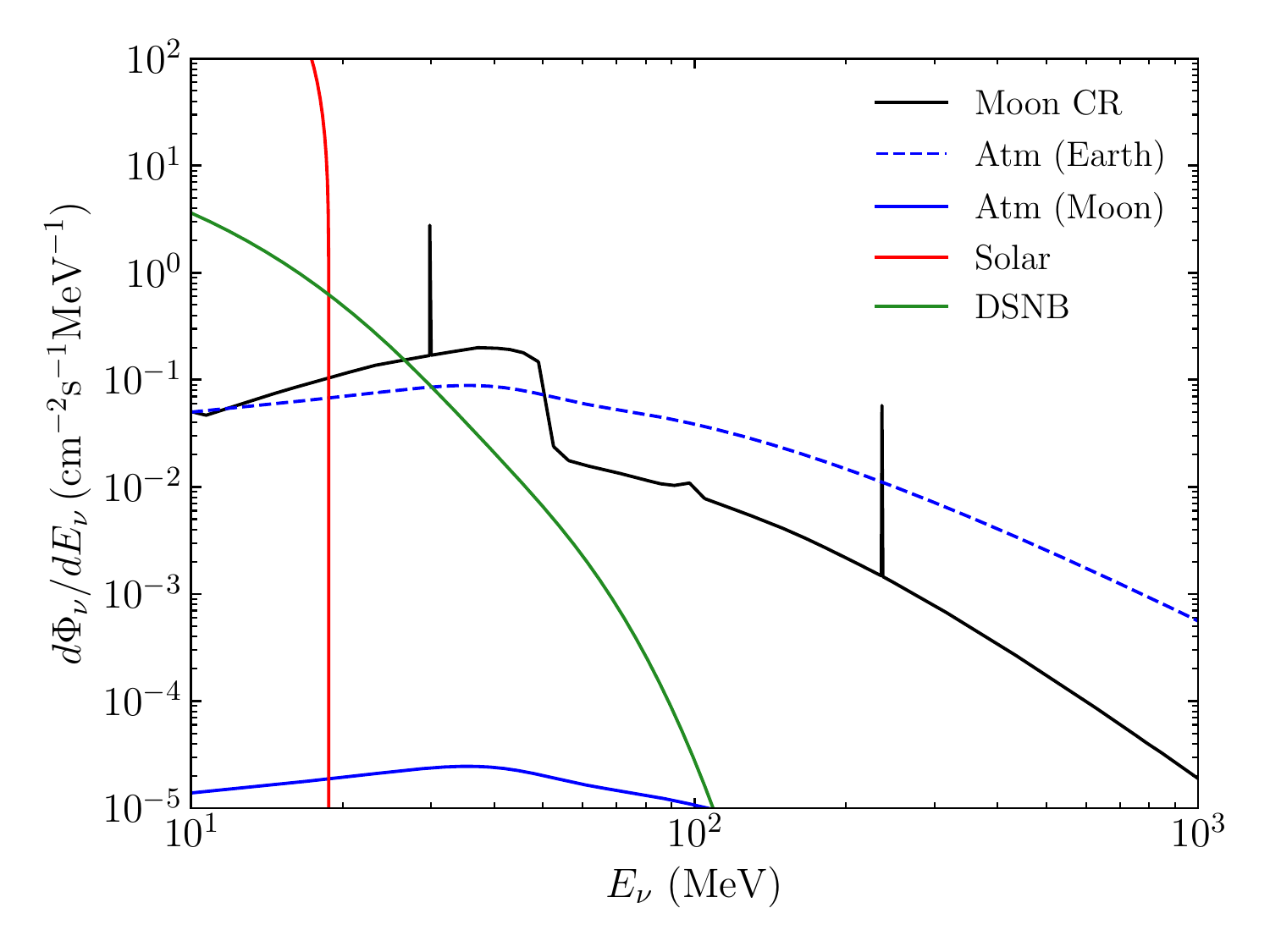}
\vspace{-0.5cm}
\caption{
  \raggedright{
    \small 
    Neutrino fluxes relevant for direct detection per unit energy, on the Moon at a depth $d=1\,\text{km}$, from cosmic ray collisions on the Moon~(black), solar~(red), DSNB~(green), and residual atmospheric neutrinos~(blue). For reference, we also show the atmospheric neutrino spectrum on the Earth~(blue dashed).
   }
      }
    \label{fig:moonflux}
\end{figure}

%\section{Lunar Neutrino Fluxes\label{sec:flux}}
{\flushleft{\bf Neutrino Fluxes:}}
Many neutrino fluxes on the Moon are nearly the same as on the Earth, but a few are radically different. For direct detection of electroweak scale dark matter on the Earth, the most important flux sources are solar neutrinos~\cite{bahcallshape,Robertson:2012ib,Vinyoles:2016djt}, diffuse supernova background neutrinos~(DSNB)~\cite{Beacom:2010kk}, and atmospheric neutrinos~\cite{Barr:1989ru}. Solar and DSNB neutrino fluxes are effectively identical on the Moon. In contrast the flux of neutrinos created by cosmic rays~(CR) -- called atmospheric neutrinos on the Earth -- differs strongly on the Moon due to its near complete lack of atmosphere~\cite{lunarsourcebook,Moondata}. Instead, CRs impact primarily on the lunar surface to produce a neutrino spectrum that is very different from that of atmospheric neutrinos on the Earth. We show the neutrino fluxes from these sources in Fig.~\ref{fig:moonflux}.

To obtain the CR neutrino flux on the Moon we make use of the results of Ref.~\cite{Demidov:2020bff}, which performed a detailed GEANT4~\cite{Brun:1994aa,GEANT4:2002zbu} simulation of the neutrino spectrum created by proton or helium cosmic rays striking the Moon. From their work we extract $dN_\nu/dE_\nu$, the mean number of neutrinos produced per cosmic ray per unit neutrino energy. This depends on the primary CR energy spectra, which are estimated through fits to the collected CR data in Ref.~\cite{Zyla:2020zbs}. From their simulations, Ref.~\cite{Demidov:2020bff} finds that pions, kaons, and muons created in cosmic ray collisions are mostly stopped in the lunar regolith and crust before decaying. Neutrinos come primarily from $\pi^+$ and $K^+$ decays and their downstream products, leading to approximately monoenergetic peaks at $E_\nu \simeq 29.8\,\mev$~($\pi^+$ decay) and $E_\nu \simeq 235.6\,\mev$~($K^+$ decay) together with a broad shoulder up to $E_\nu \simeq 52.8\,\mev$~(3-body $\mu^+$ decay). In our analysis, we model the $\pi^+$ and $K^+$ peaks with delta functions in energy. The Moon CR neutrinos are concentrated at lower energies than atmospheric neutrinos, which are  mostly created by decays in flight~\cite{Barr:1989ru}. For the neutrino energies of primary interest, $E_\nu \lesssim 1\,\gev$, the production is found to occur mainly within the first tens of meters of the surface~\cite{Demidov:2020bff}. We have verified this independently with simulations in GEANT4~\cite{GEANT4:2002zbu} and FLUKA~\cite{Ferrari:2005zk}.

We are interested in the flux of neutrinos at a depth $d$ below the surface of the Moon. We take this depth to be considerably larger than the thickness of the area over which the neutrinos are produced. This allows us to treat the emission as coming from a shell at the surface. As in Ref.~\cite{Demidov:2020bff}, we also treat the emission of neutrinos as being isotropic. With these assumptions, the flux of neutrinos per unit energy at depth $d$ is:
\beq
\frac{d\Phi_\nu}{dE_\nu} = \frac{dN_\nu}{dE_\nu}\,{\Phi_{\rm CR}}\,\frac{r_M}{8(r_M-d)}\,\ln\lrf{2r_M-d}{d} \ ,
\label{eq:mcrflux}
\eeq
where $\Phi_{\rm CR}$ is the total CR flux striking the Moon integrated over all angles and CR energies and subject to the energy cuts made in Ref.~\cite{Demidov:2020bff} when computing $dN_\nu/dE_\nu$ (kinetic energy $E_k > 400\,\mev$ for protons and $E_k > 800\,\mev$ for helium), and $r_M\simeq 1740\,\text{km}$ is the radius of the Moon. For the analysis to follow we fix $d=1\,\text{km}$ as a representative value of the depth of the detector. Small variations in the depth (by a factor of a few) do not significantly change our results.

\begin{comment}
We are interested in the flux of neutrinos at a depth $d$ below the surface of the Moon considerably larger than the thickness over which they are produced. This allows us to treat the emission as coming from a shell at the surface. As in Ref.~\cite{Demidov:2020bff}, we also treat the surface emission of neutrinos as being isotropic. With these assumptions, the total rate of neutrino emission per unit area of the lunar surface is 
\beq
\frac{dJ_\nu}{dE_\nu} = \frac{1}{4}\,\Phi_{\rm CR}\,\frac{dN_\nu}{dE_\nu} \ ,
\label{eq:mcrnu}
\eeq
where $\Phi_{\rm CR}$ is the total CR flux striking the Moon integrated over solid angle and energy and subject to the energy cuts made in Ref.~\cite{Demidov:2020bff} when computing $N_\nu$ ($E_k > 400\,\mev$ for protons and $E_k > 800\,\mev$ for helium). The factor of $1/4$ comes from integrating the normal component of the CR flux over half its total range to account for only downward-going cosmic rays contributing~\cite{Demidov:2020bff}.  Summing over the entire surface, the lunar CR neutrino flux at a depth $d$ below it is
\beq
\frac{d\Phi_\nu}{dE_\nu} = \frac{dJ_\nu}{dE_\nu}\,\frac{r_M}{2(r_M-d)}\,\ln\lrf{2r_M-d}{d} \ ,
\label{eq:mcrflux}
\eeq
where $r_M\simeq 1740\,\text{km}$ is the radius of the Moon. For the analysis to follow, we set $d=1\,\text{km}$ as a fiducial value.
\end{comment}

 This calculation of the CR neutrino flux is expected to provide a sufficiently good approximation to estimate the sensitivity of DM detectors on the Moon. However, we note that it could be refined in several ways that go beyond the scope of this letter. The assumption of isotropic neutrino emission from the lunar surface is expected to hold for neutrinos from decays at rest, but for the rarer decays in flight there can be a correlation between the incident CR direction and the angular spectrum of the neutrinos produced. As shown in the \hyperref[sec:sup]{Supplemental Material}, we find that for uniform neutrino emission directed within a finite cone along the CR direction, the flux at $d=1\,\text{km}$ varies by less than about $30\%$ for any cone angle, and thus we expect that going beyond the isotropic approximation would not significantly change our results. There is also an uncertainty in the composition and density of the lunar regolith and crust. The calculation of Ref.~\cite{Demidov:2020bff} suggests that the impact of this uncertainty on the neutrino flux is modest, and we note that it could be reduced with future exploration of the Moon. Finally, the CR flux with $E_{\rm CR}\lesssim 10\,\gev$ varies over the solar cycle~\cite{Gleeson:1968zza,Potgieter:2013pdj} with the value used here representing a mean. This uncertainty could be largely eliminated by monitoring the primary cosmic ray spectra with a modest CR detector on the lunar surface.

Other potential sources of neutrinos on the Moon are radioactive decays of lunar material, atmospheric neutrinos from the Earth, and solar energetic particles. All three can be neglected in this analysis to a good approximation. Neutrinos from radioactive decays are analogous to geoneutrinos in the Earth, have energies below $E_\nu \lesssim 10\,\mev$, and are negligible in comparison to solar neutrinos which dominate in this energy range~\cite{Gelmini:2018gqa}. Atmospheric neutrinos from the Earth are suppressed by the factor $(r_\oplus/R)^2 \simeq 3\times 10^{-4}$, where $r_\oplus \simeq 6400\,\text{km}$ is the Earth radius and $R \simeq 384400\,\text{km}$ is the Moon distance, and their flux is much smaller than Moon CR neutrinos. Solar energetic particles~(SEP) are emitted sporadically in solar events and can sometimes include protons and helium with enough energy to create pions~\cite{Reames:2020vxe,PAMELA:2018yie}. These are present for only a small fraction of the observing time and their effects can be removed by discarding data taken during these periods. 

In Fig.~\ref{fig:moonflux} we show the energy spectra of the most significant neutrino fluxes on the Moon summed over flavors and including anti-neutrinos. The relevant sources are solar, DSNB, and cosmic rays striking the Moon. For reference we also show the atmospheric neutrino spectrum on the Earth as well as its residual value on the Moon. Note that the heights of the pion and kaon decay lines in the figure correspond to the normalizations of the delta functions used to represent them. Relative to the atmospheric spectrum on the Earth, we see that the CR neutrino spectrum on the Moon is shifted to lower energies and is concentrated at specific locations.

%\section{Dark Matter and Neutrino Scattering}
{\flushleft{\bf Dark Matter and Neutrino Scattering:}}
The primary search objective of large-scale noble-liquid dark matter detectors is the scattering of a dark matter species $\chi$ off nuclei. In a detector target composed of nucleus $N=(A,Z)$, this leads to recoil energy depositions with a rate per unit target mass of:
\beq
\frac{d\widetilde{R}_{\chi}^{(N)}}{dE_R} = 
{\varepsilon_{NR}}\,{n_N}\lrf{\rho_\chi}{m_\chi}\int_{v_{min}}\!\!d^3v\,f(\vec{v})\,v\,\frac{d\sigma_{\chi N}}{dE_R} \ ,
\eeq
where $E_R$ is the nuclear recoil~(NR) energy, $\varepsilon_{NR}$ is the NR detection efficiency, $n_N$ is the number of $N$ nuclei per unit mass, $\rho_\chi$ is the DM mass density and $m_\chi$ is the DM particle mass, $f(\vec{v})$ is the local DM velocity distribution restricted to $\lVert \vec{v}\rVert > v_{min} \equiv \sqrt{m_NE_R/2\mu_N^2}$ for DM-nucleus reduced mass $\mu_N$, and $d\sigma_{\chi N}/dE_R$ is the DM-nucleus differential cross section. For primarily spin-independent~(SI) DM-nucleus scattering, the full cross section can be expressed in terms of an effective DM-nucleon cross section ${\sigma}_n$ as described in Ref.~\cite{Jungman:1995df} and the \hyperref[sec:sup]{Supplemental Material}.

\begin{figure*}[t!]
    \centering
\makebox[\textwidth][c]{  \includegraphics[width=.47\textwidth]{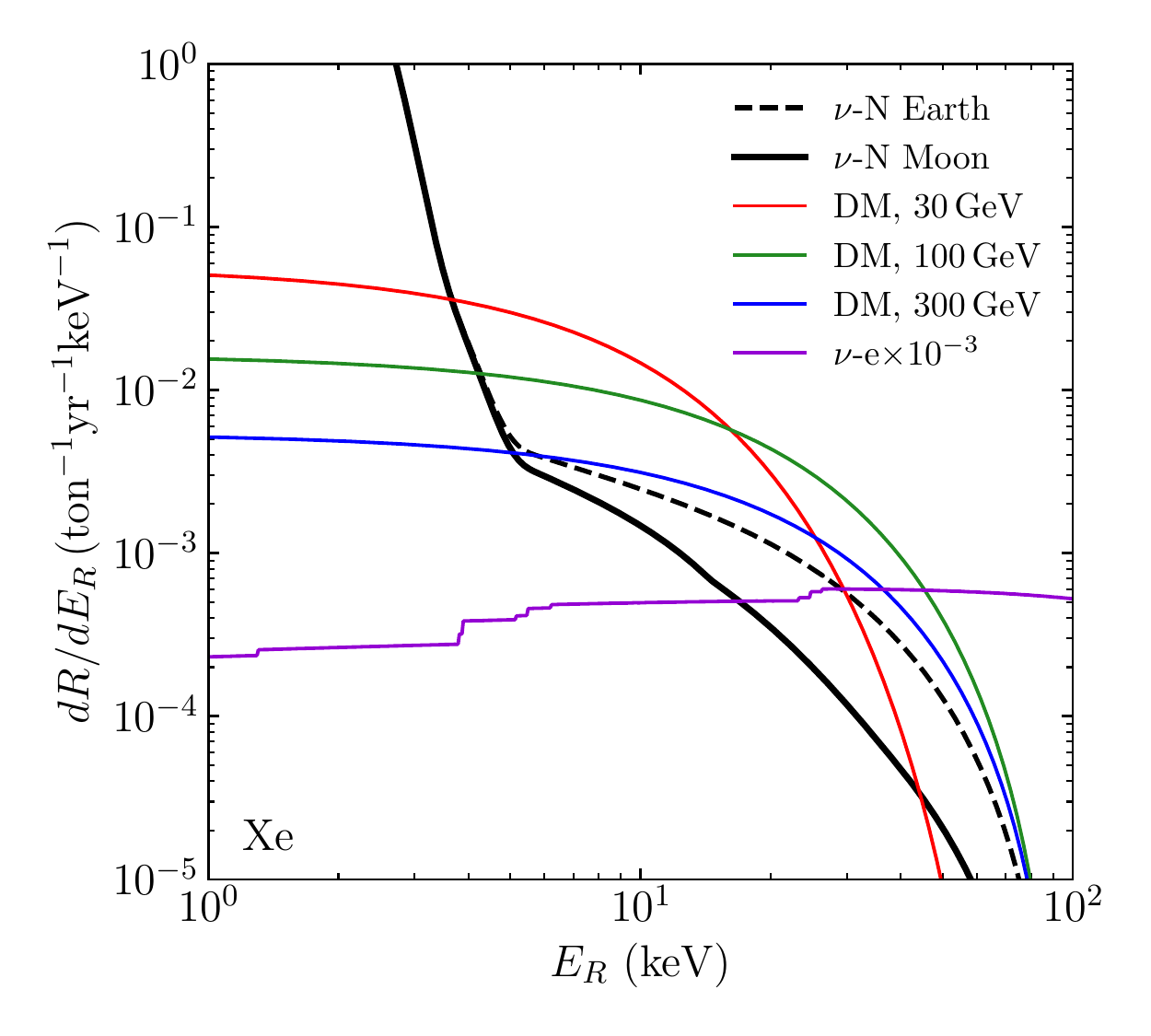} \
 \includegraphics[width=.47\textwidth]{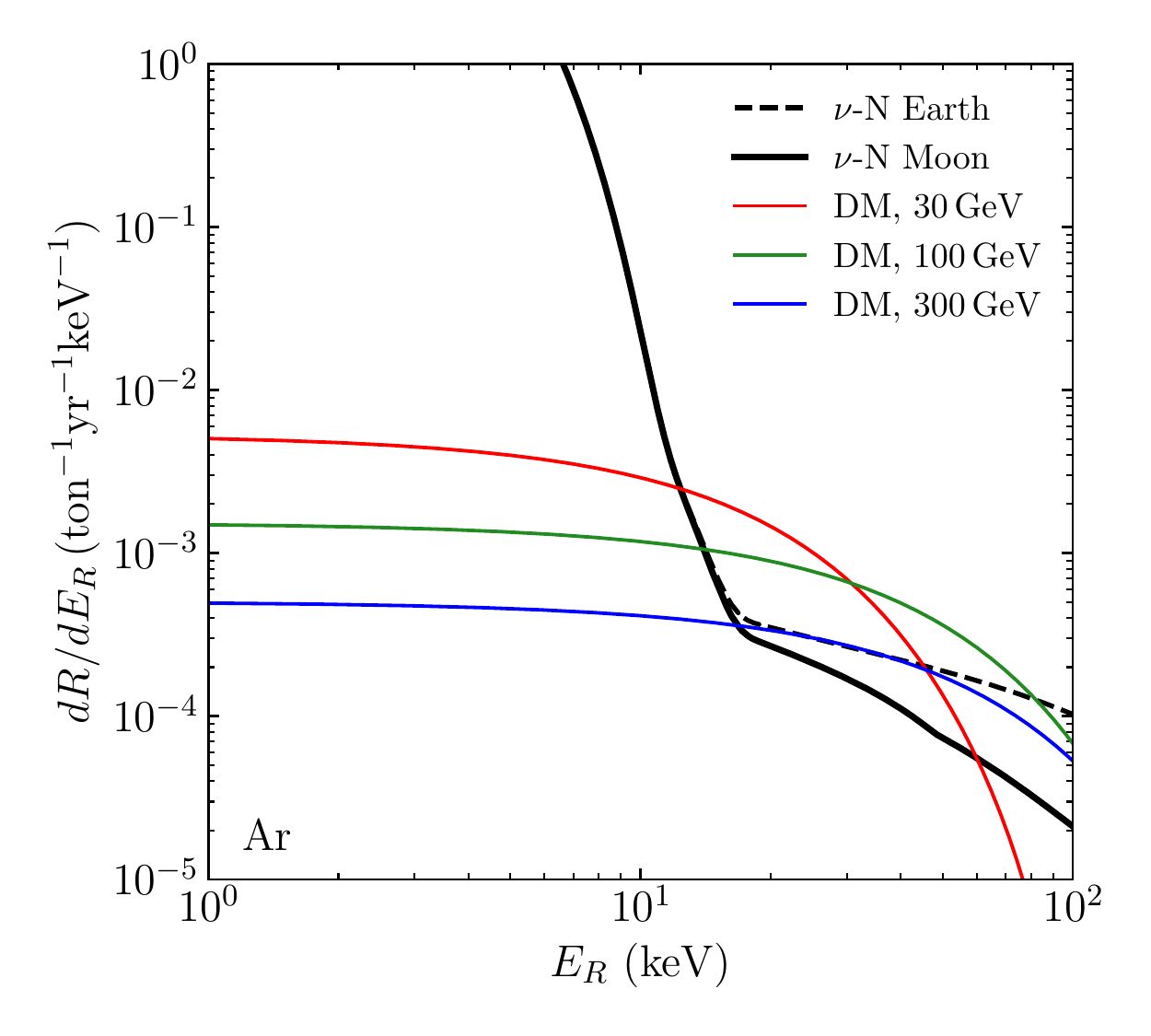} 
 }
 \vspace{-0.5cm}
\caption{
  \raggedright{
    \small 
    Recoil rates per unit target mass as a function of nuclear recoil energy $E_R$ in xenon~(left) and argon~(right) from background neutrinos on the Moon~(black solid), on the Earth~(black dashed), and dark matter with $\sigma_n=10^{-48}\,\text{cm}^2$ and $m_\chi = 30,\,100,\,300\,\gev$~(red, green, blue). For xenon, we also show the rate for misidentified neutrino-electron scattering, scaled by a representative electron rejection factor of $10^{-3}$.
   }
      }
    \label{fig:erspectra}
\end{figure*}

Neutrino scattering on nuclei leads to a similar expression for the rate per unit target mass,
\beq
\frac{d\widetilde{R}_\nu^{(N)}}{dE_R} = {\varepsilon_{NR}}\,{n_N}\,\int_{E_{min}}\!\!\!dE_\nu\,\sum_{j}\frac{d\Phi_j}{dE_\nu}\,\frac{d\sigma_{\nu N}}{dE_R} \ ,
\eeq
with $E_{min} = \sqrt{m_NE_R/2}$ and the sum on $j$ runs over relevant neutrino flux sources $\Phi_j(E_\nu)$. Note that these fluxes are summed over all neutrino flavors and implicitly include antineutrinos since the underlying scattering process is flavor-independent $Z^0$ boson exchange. A full expression for $d\sigma_{\nu N}/dE_R$ is given in Ref.~\cite{Freedman:1977xn} and the \hyperref[sec:sup]{Supplemental Material}.

In realistic detectors, neutrino scattering on atomic electrons can be mistaken for scattering on nuclei with probability $\varepsilon_e(E_R)$. When this occurs, the average reconstructed nuclear recoil energy $E_R$ can be related to the actual energy transfer $T$ by a factor $q_{eff}$~\cite{Szydagis:2021hfh},
\beq
T = q_{eff}(E_R)\;E_R \ . 
\eeq
The contribution of neutrino-electron scattering to the reconstructed nuclear recoil rate is therefore
\beq
\frac{d\widetilde{R}_\nu^{(N,Ze)}}{dE_R} = \varepsilon_e(E_R)\left[\frac{dT}{dE_R}\frac{d\widetilde{R}_\nu^{(Ze)}}{dT}\right]_{T=q_{eff}E_R} \ ,
\eeq
with
\beq
\frac{d\widetilde{R}_\nu^{(Ze)}}{dT} = {n_N}\,\int_{E_{min}}\!\!\!dE_\nu\,\sum_{j,a}\frac{d\Phi_{j,a}}{dE_\nu}\,\frac{d\sigma_{a}^{(Ze)}}{dT~} 
\eeq
where $E_{min} = [T+\sqrt{T(T+2m_e)}]$ and we have specified the neutrino flavor $a$ since scattering on atomic electrons is flavor-dependent. Explicit expressions for $d\sigma_a^{(e)}/dT$ for scattering on free electrons can be found in Refs.~\cite{Sarantakos:1982bp,Vogel:1989iv,Marciano:2003eq} and the \hyperref[sec:sup]{Supplemental Material}, while detailed \emph{ab initio} calculations for scattering on atomic electrons in xenon are performed in Ref.~\cite{Chen:2016eab}, the latter of which are used in our analysis.

To evaluate these rates, we also need estimates for the local DM distribution and the relevant neutrino fluxes. For DM, we follow the recommendations of Ref.~\cite{Baxter:2021pqo} and use $\rho_\chi = 0.3\,\gev/\text{cm}^3$ with the Standard Halo Model velocity distribution with parameters $v_0 = 238\,\text{km/s}$, $v_E = 254\,\text{km/s}$, and $v_{esc} = 544\,\text{km/s}$. Neutrino fluxes and their uncertainties are taken as in Ref.~\cite{Gaspert:2021gyj}, which mostly follows Ref.~\cite{Baxter:2021pqo}. For each flux source, we model the differential spectrum as
\beq
\frac{d\Phi_j}{dE_\nu} = \phi_j\,f_j(E_\nu) \ ,
\label{eq:fluxfunc}
\eeq
where $j=1,2,\ldots,n_\nu$ runs over flux sources, $\phi_j$ is the total flux, and the spectral functions $f_j(E_\nu)$ are normalized to unity. Solar neutrino fluxes in this work are based on the high metallicity model~\cite{Grevesse:1998bj,Asplund:2009fu,Serenelli:2011py} with specific predictions from Refs.~\cite{Baxter:2021pqo,Vinyoles:2016djt} and updated with recent data~\cite{Bergstrom:2016cbh,Agostini:2020mfq}. For the total atmospheric neutrino flux on the Earth, we use the results of Ref.~\cite{Battistoni:2005pd} relevant for a detector located at Gran Sasso and averaged over the solar cycle with an estimated fractional uncertainty of 20\%~\cite{Battistoni:2005pd,Barr:2006it,Honda:2011nf,Honda:2015fha}. We also estimate a fractional uncertainty of 20\% for CR neutrino fluxes on the Moon, although we expect that this could be improved with direct measurements of the primary CR spectrum.

In Fig.~\ref{fig:erspectra} we show the scattering rates for dark matter and neutrinos in xenon~(left) and argon~(right) as functions of the reconstructed nuclear recoil energy $E_R$ for $\varepsilon_{NR}=1$. The red, green, and blue lines show rates from DM scattering with $\sigma_n = 10^{-48}\,\text{cm}^2$ for masses $m_\chi = 30,\,100,\,300\,\gev$ respectively. The solid black line in each panel shows the total neutrino-nucleus recoil rate spectrum on the Moon, while the dashed black line shows the corresponding spectrum on the Earth. These are significantly different, with lower rate at $E_R \gtrsim 5\,\kev$~($20\,\kev$) for xenon (argon) on the Moon. The solid purple line in the left panel shows the reconstructed spectrum from neutrino-electron scatterings misidentified as nuclear scatterings and rescaled by $10^{-3}$ to represent a typical electron rejection factor in xenon. No such line is shown for argon where electron rejection is expected to be much more efficient.

%\section{Dark Matter Sensitivity Estimates}
{\flushleft{\bf Dark Matter Sensitivity:}}
We turn next to estimating the dark matter discovery sensitivity of detectors located on the Moon. Following Ref.~\cite{Gaspert:2021gyj}, we focus on large-scale xenon and argon detectors with properties based on current and proposed experiments. We also treat statistical and systematic uncertainties using a binned profile likelihood evaluated in the asymptotic limit~\cite{Cowan:2010js,Tang:2023xub}

\begin{figure*}[ttt]
    \centering
\makebox[\textwidth][c]{  \includegraphics[width=.47\textwidth]{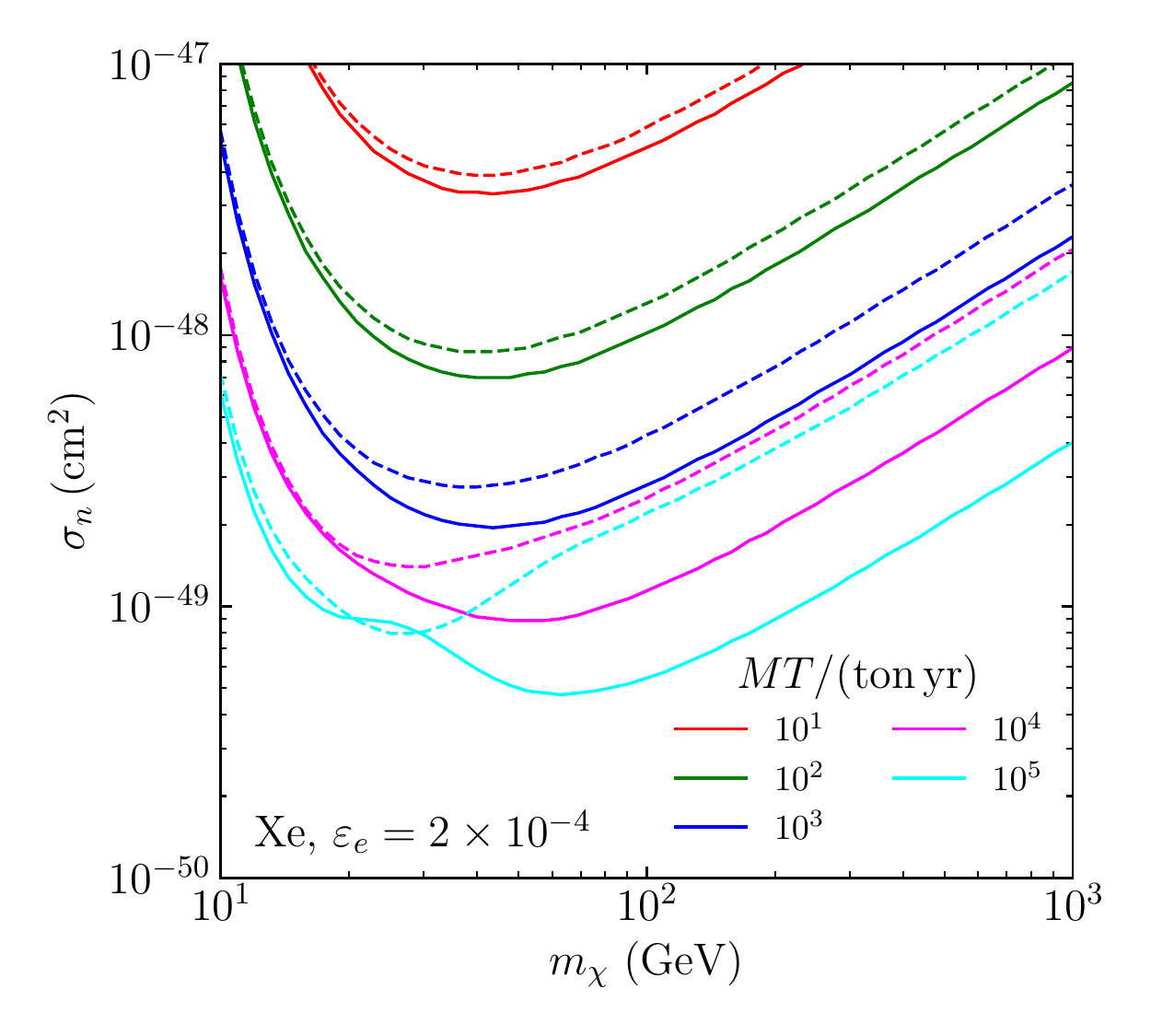} \
 \includegraphics[width=.47\textwidth]{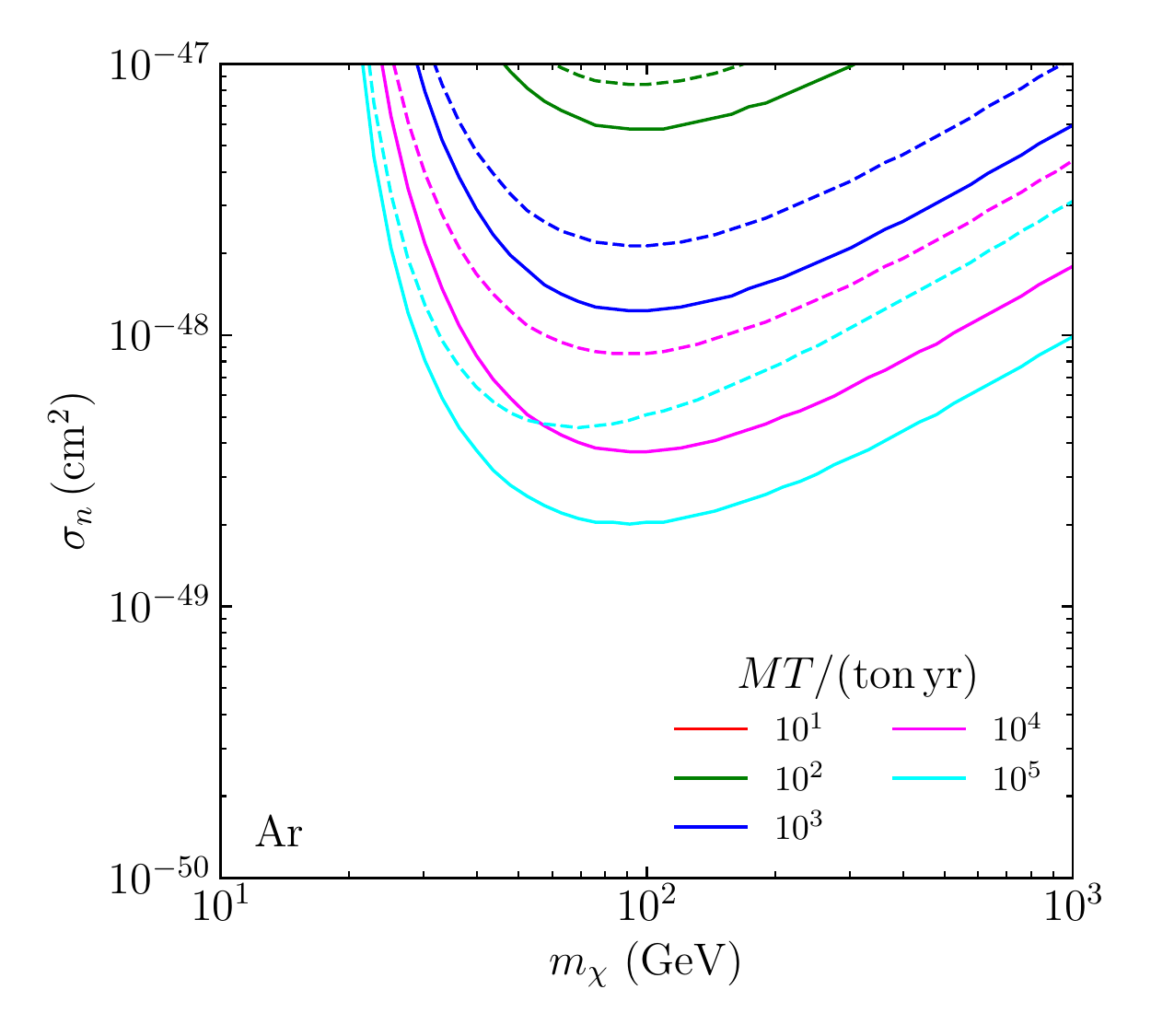} 
 }
 \vspace{-0.5cm}
\caption{
  \raggedright{
    \small 
    Sensitivity to SI dark matter with per nucleon cross section $\sigma_n$ as a function of DM mass $m_\chi$ for several values of detector exposure $MT/(\text{ton\,yr})$ in representative xenon~(left) and argon~(right) detectors. The solid lines indicate the sensitivities for detectors on the Moon and the dashed lines for detectors on the Earth.
   }
      }
    \label{fig:nufloor1}
\end{figure*}

For xenon detectors, we model their parameters on the planned DARWIN experiment~\cite{DARWIN:2016hyl,Aalbers:2022dzr}. We assume an analysis region of interest~(ROI) for nuclear recoil energies $E_R\in [5,35]\,\kev$ and we collect events over this energy region into bins with size dictated by the local energy resolution as described in Ref.~\cite{Gaspert:2021gyj}. We also take a constant nuclear recoil efficiency of $\varepsilon_{NR}=0.5$ and an electron recoil rejection factor of $\varepsilon_e = 2\times 10^{-4}$.

In argon, we model detector properties on the proposed ARGO experiment~\cite{Galbiati:2018} with guidance from DEAP-3600~\cite{DEAP:2019yzn} and DarkSide~\cite{Aalseth:2017fik}. We use a nuclear recoil ROI of $E_R\in [55,100]\,\kev$ with $\varepsilon_{NR}=0.9$. As for xenon, we collect events in bins dictated by the expected local energy resolution in argon~\cite{Gaspert:2021gyj}. Pulse-shape discrimination allows for excellent rejection of electron recoils~\cite{Boulay:2006mb,DEAP:2009hyz,Adhikari:2020zyy,DEAP:2021axq}, and we take $\varepsilon_e < 10^{-8}$.

To estimate the DM discovery potential of a lunar detector, we use the profile likelihood method~\cite{Cowan:2010js} For a given DM mass $m_\chi$, per-nucleon cross section $\sigma_n$, and detector configuration and exposure, we classify the scenario as discoverable if the background-only hypothesis (\emph{i.e.} neutrinos only) can be excluded at the $3\sigma$ level %($p<0.0013$) 
at least 90\% of the time. We arrange events into recoil energy bins $N^i$, $i=1,2,\ldots, N_b$, with the expected number of events in each bin given by
\beq
\langle N^i\rangle = \xi\,s^i+b^i(\vec{\theta}) \ ,
\eeq
where $\xi = \sigma_n/\sigma_0$ is the DM cross section relative to an arbitrary reference value, $s^i$ is the number of DM events with $\sigma_0$, $b^i(\vec{\theta})$ is the number of background events, and $\vec{\theta}$ is a set of parameters that describe the backgrounds. We identify them with fractional neutrino flux variations,
\beq
\theta_j = \frac{\phi_j-\bar{\phi}_j}{\bar{\phi}_j}
\eeq
where $\phi_j$ is the flux normalization of Eq.~\eqref{eq:fluxfunc} for source $j$ relative to the central value $\bar{\phi}_j$. The likelihood used is
\beq
\mathcal{L}(\xi,\vec{\theta}) = \prod_{i=1}^{N_b}\mathcal{P}(N^i,\langle N^i\rangle)\times \prod_{j=1} ^{n_\nu}\frac{e^{-(\theta_j/\Delta\theta_j)^2/2}}{\sqrt{2\pi}\,\Delta\theta_j}  
\eeq
with $\mathcal{P}(\lambda,n) = \lambda^ne^{-\lambda}/n!$ and $\Delta\theta_j$ is the fractional flux uncertainty. In the asymptotic limit of many events, the results of Ref.~\cite{Cowan:2010js} based on Wilks' theorem~\cite{Wilks:1938dza,Wald:1942} can be extended to show that our DM discovery condition is equivalent to the test statistic $q_{0,A} > 18.34$~\cite{Gaspert:2021gyj}, where 
\beq
q_{0,A} = -2\ln\!\left[\mathcal{L}(\xi=0,\widehat{\vec{\theta}})\Big/\mathcal{L}(\xi,\vec{\theta}=\vec{0})\right] \ .
\eeq
Here, $\widehat{\vec{\theta}}$ are the values of the background parameters that optimize the likelihood function under the background-only hypothesis $\xi = 0$ for the dataset $N^i = \langle N^i\rangle$ obtained with DM signal $\xi$ and neutrino fluxes $\vec{\theta}=\vec{0}$. 

With these methods we obtain the dark matter sensitivities to the per-nucleon SI cross section $\sigma_n$ shown in Fig.~\ref{fig:nufloor1} for representative xenon~(left) and argon~(right) detectors as a function of DM mass $m_\chi$.  The solid lines indicate the sensitivities for a detector on the Moon and the dashed lines show sensitivities on the Earth. Lines of different colors correspond to the indicated total exposures $MT/(\text{ton\,yr})=10^2,\,10^3,\,10^4,\,10^5$~(top to bottom).

The contours in Fig.~\ref{fig:nufloor1} demonstrate that an equivalent detector located on the Moon could have better sensitivity to DM than a detector on the Earth. This is due to the very different neutrino flux spectra induced by cosmic rays in the two locations, as shown in Fig.~\ref{fig:moonflux}. The sensitivity improvement on the Moon is seen primarily for masses $m_\chi \gtrsim 50\,\gev$ and is more pronounced for xenon than argon. In xenon, the recoil energy spectrum from atmospheric neutrinos is closer to that from dark matter compared to argon, and in this sense the neutrino fog is more opaque for xenon detectors. The modified CR neutrino spectrum on the Moon reduces this spectral degeneracy and the total background rate, leading to sensitivity improvements in xenon by up to nearly an order of magnitude in $\sigma_n$. A greater DM sensitivity is also seen in argon but the improvement is more modest. We also find that these results are relatively robust under reasonable variations in the electron rejection factor and in the uncertainty in the Moon CR neutrino flux as shown in the \hyperref[sec:sup]{Supplemental Material}.

%\section{Summary and Conclusions} \label{sec:concs}}
%%%%%
\flushleft{\bf Summary and Conclusions:}
In this work we have estimated the fluxes of neutrinos below the surface of the Moon by leveraging the simulations of Ref.~\cite{Demidov:2020bff}. The spectrum of neutrinos created by cosmic rays is found to be significantly different on the Moon. We applied these results to calculate the potential sensitivities of large-scale xenon and argon detectors located on the Moon to spin-independent dark matter~(DM). Neutrino backgrounds from cosmic ray collisions are easier to distinguish from dark matter on the Moon relative to the Earth. All else being equal, this implies that a DM detector located on the Moon would have a significantly greater sensitivity to DM with mass $m_\chi \gtrsim 50\,\gev$ than one on the Earth. In particular, this could enable sensitivity to the $m_\chi = 1.1\,\tev$ thermal Higgsino DM~\cite{Hill:2013hoa,Hill:2014yxa,Co:2021ion,Dessert:2022evk}.

While our results provide motivation for Moon-based DM direct detection experiments, we have not attempted to address the many practical challenges to realizing such an apparatus. These include transportation and cosmic activation, local backgrounds, infrastructure, and human resources. Even so, given the current drive to return to the Moon~\cite{artemis}, it is worthwhile and timely to consider this and other potential applications of lunar exploration to scientific discovery.

%\section*{Acknowledgments}
\flushleft{\bf Acknowledgments:}
We thank Alan Robinson for pointing out that cosmic ray induced neutrino fluxes would be different on the Moon relative to the Earth. We also thank Nikita Blinov, Dave McKeen, Marianne Moore, Scott Oser, Nirmal Raj, and Michael Shamma for helpful discussions.
This work is supported in part by the Natural Sciences and Engineering Research Council (NSERC) of Canada. TRIUMF receives federal funding via a contribution agreement with the National Research Council (NRC) of Canada.

\bibliography{refs.bib}

\begin{thebibliography}{102}%
\makeatletter
\providecommand \@ifxundefined [1]{%
 \@ifx{#1\undefined}
}%
\providecommand \@ifnum [1]{%
 \ifnum #1\expandafter \@firstoftwo
 \else \expandafter \@secondoftwo
 \fi
}%
\providecommand \@ifx [1]{%
 \ifx #1\expandafter \@firstoftwo
 \else \expandafter \@secondoftwo
 \fi
}%
\providecommand \natexlab [1]{#1}%
\providecommand \enquote  [1]{``#1''}%
\providecommand \bibnamefont  [1]{#1}%
\providecommand \bibfnamefont [1]{#1}%
\providecommand \citenamefont [1]{#1}%
\providecommand \href@noop [0]{\@secondoftwo}%
\providecommand \href [0]{\begingroup \@sanitize@url \@href}%
\providecommand \@href[1]{\@@startlink{#1}\@@href}%
\providecommand \@@href[1]{\endgroup#1\@@endlink}%
\providecommand \@sanitize@url [0]{\catcode `\\12\catcode `\$12\catcode
  `\&12\catcode `\#12\catcode `\^12\catcode `\_12\catcode `\%12\relax}%
\providecommand \@@startlink[1]{}%
\providecommand \@@endlink[0]{}%
\providecommand \url  [0]{\begingroup\@sanitize@url \@url }%
\providecommand \@url [1]{\endgroup\@href {#1}{\urlprefix }}%
\providecommand \urlprefix  [0]{URL }%
\providecommand \Eprint [0]{\href }%
\providecommand \doibase [0]{http://dx.doi.org/}%
\providecommand \selectlanguage [0]{\@gobble}%
\providecommand \bibinfo  [0]{\@secondoftwo}%
\providecommand \bibfield  [0]{\@secondoftwo}%
\providecommand \translation [1]{[#1]}%
\providecommand \BibitemOpen [0]{}%
\providecommand \bibitemStop [0]{}%
\providecommand \bibitemNoStop [0]{.\EOS\space}%
\providecommand \EOS [0]{\spacefactor3000\relax}%
\providecommand \BibitemShut  [1]{\csname bibitem#1\endcsname}%
\let\auto@bib@innerbib\@empty
%</preamble>
\bibitem [{\citenamefont {Ruggles}(2015)}]{moon2015}%
  \BibitemOpen
  \bibinfo {editor} {\bibfnamefont {C.~L.}\ \bibnamefont {Ruggles}},\ ed.,\
  \href {\doibase 10.1007/978-1-4614-6141-8} {\emph {\bibinfo {title} {Handbook
  of Archaeoastronomy and Ethnoastronomy}}}\ (\bibinfo  {publisher} {Springer
  New York},\ \bibinfo {year} {2015})\BibitemShut {NoStop}%
\bibitem [{\citenamefont {{Heiken}}\ \emph {et~al.}(1991)\citenamefont
  {{Heiken}}, \citenamefont {{Vaniman}},\ and\ \citenamefont
  {{French}}}]{lunarsourcebook}%
  \BibitemOpen
  \bibfield  {author} {\bibinfo {author} {\bibfnamefont {G.~H.}\ \bibnamefont
  {{Heiken}}}, \bibinfo {author} {\bibfnamefont {D.~T.}\ \bibnamefont
  {{Vaniman}}}, \ and\ \bibinfo {author} {\bibfnamefont {B.~M.}\ \bibnamefont
  {{French}}},\ }\href@noop {} {\emph {\bibinfo {title} {{Lunar Sourcebook, A
  User's Guide to the Moon}}}}\ (\bibinfo {year} {1991})\BibitemShut {NoStop}%
\bibitem [{\citenamefont {Dickey}\ \emph {et~al.}(1994)\citenamefont {Dickey}
  \emph {et~al.}}]{Dickey:1994zz}%
  \BibitemOpen
  \bibfield  {author} {\bibinfo {author} {\bibfnamefont {J.~O.}\ \bibnamefont
  {Dickey}} \emph {et~al.},\ }\href {\doibase 10.1126/science.265.5171.482}
  {\bibfield  {journal} {\bibinfo  {journal} {Science}\ }\textbf {\bibinfo
  {volume} {265}},\ \bibinfo {pages} {482} (\bibinfo {year}
  {1994})}\BibitemShut {NoStop}%
\bibitem [{\citenamefont {Williams}\ \emph {et~al.}(2004)\citenamefont
  {Williams}, \citenamefont {Turyshev},\ and\ \citenamefont
  {Boggs}}]{Williams:2004qba}%
  \BibitemOpen
  \bibfield  {author} {\bibinfo {author} {\bibfnamefont {J.~G.}\ \bibnamefont
  {Williams}}, \bibinfo {author} {\bibfnamefont {S.~G.}\ \bibnamefont
  {Turyshev}}, \ and\ \bibinfo {author} {\bibfnamefont {D.~H.}\ \bibnamefont
  {Boggs}},\ }\href {\doibase 10.1103/PhysRevLett.93.261101} {\bibfield
  {journal} {\bibinfo  {journal} {Phys. Rev. Lett.}\ }\textbf {\bibinfo
  {volume} {93}},\ \bibinfo {pages} {261101} (\bibinfo {year} {2004})},\
  \Eprint {http://arxiv.org/abs/gr-qc/0411113} {arXiv:gr-qc/0411113}
  \BibitemShut {NoStop}%
\bibitem [{\citenamefont {Will}(2014)}]{Will:2014kxa}%
  \BibitemOpen
  \bibfield  {author} {\bibinfo {author} {\bibfnamefont {C.~M.}\ \bibnamefont
  {Will}},\ }\href {\doibase 10.12942/lrr-2014-4} {\bibfield  {journal}
  {\bibinfo  {journal} {Living Rev. Rel.}\ }\textbf {\bibinfo {volume} {17}},\
  \bibinfo {pages} {4} (\bibinfo {year} {2014})},\ \Eprint
  {http://arxiv.org/abs/1403.7377} {arXiv:1403.7377 [gr-qc]} \BibitemShut
  {NoStop}%
\bibitem [{\citenamefont {NASA}(2019)}]{artemis}%
  \BibitemOpen
  \bibfield  {author} {\bibinfo {author} {\bibnamefont {NASA}},\ }\href
  {https://www.nasa.gov/artemisprogram} {\enquote {\bibinfo {title} {Artemis
  program},}\ }\bibinfo {howpublished}
  {\url{https://www.nasa.gov/artemisprogram}} (\bibinfo {year} {2019}),\
  \bibinfo {note} {accessed: 2023-04-20}\BibitemShut {NoStop}%
\bibitem [{\citenamefont {Silk}(2021)}]{Silk:2020bsr}%
  \BibitemOpen
  \bibfield  {author} {\bibinfo {author} {\bibfnamefont {J.}~\bibnamefont
  {Silk}},\ }\href {\doibase 10.1098/rsta.2019.0561} {\bibfield  {journal}
  {\bibinfo  {journal} {Phil. Trans. A. Math. Phys. Eng. Sci.}\ }\textbf
  {\bibinfo {volume} {379}},\ \bibinfo {pages} {20190561} (\bibinfo {year}
  {2021})},\ \Eprint {http://arxiv.org/abs/2011.04671} {arXiv:2011.04671
  [astro-ph.CO]} \BibitemShut {NoStop}%
\bibitem [{\citenamefont {Valinia}\ \emph {et~al.}(2022)\citenamefont {Valinia}
  \emph {et~al.}}]{artemissci}%
  \BibitemOpen
  \bibfield  {author} {\bibinfo {author} {\bibfnamefont {A.}~\bibnamefont
  {Valinia}} \emph {et~al.},\ }\href
  {https://ntrs.nasa.gov/citations/20220017053} {\emph {\bibinfo {title}
  {Unique Science from the Moon in the Artemis Era}}},\ \bibinfo {type} {Tech.
  Rep.}\ \bibinfo {number} {NASA/TM-20220017053}\ (\bibinfo  {institution}
  {NASA},\ \bibinfo {address} {Washington, DC},\ \bibinfo {year}
  {2022})\BibitemShut {NoStop}%
\bibitem [{\citenamefont {Beacham}\ and\ \citenamefont
  {Zimmermann}(2022)}]{Beacham:2021lgt}%
  \BibitemOpen
  \bibfield  {author} {\bibinfo {author} {\bibfnamefont {J.}~\bibnamefont
  {Beacham}}\ and\ \bibinfo {author} {\bibfnamefont {F.}~\bibnamefont
  {Zimmermann}},\ }\href {\doibase 10.1088/1367-2630/ac4921} {\bibfield
  {journal} {\bibinfo  {journal} {New J. Phys.}\ }\textbf {\bibinfo {volume}
  {24}},\ \bibinfo {pages} {023029} (\bibinfo {year} {2022})},\ \Eprint
  {http://arxiv.org/abs/2106.02048} {arXiv:2106.02048 [hep-ex]} \BibitemShut
  {NoStop}%
\bibitem [{\citenamefont {Aghanim}\ \emph {et~al.}(2020)\citenamefont {Aghanim}
  \emph {et~al.}}]{Planck:2018vyg}%
  \BibitemOpen
  \bibfield  {author} {\bibinfo {author} {\bibfnamefont {N.}~\bibnamefont
  {Aghanim}} \emph {et~al.} (\bibinfo {collaboration} {Planck}),\ }\href
  {\doibase 10.1051/0004-6361/201833910} {\bibfield  {journal} {\bibinfo
  {journal} {Astron. Astrophys.}\ }\textbf {\bibinfo {volume} {641}},\ \bibinfo
  {pages} {A6} (\bibinfo {year} {2020})},\ \bibinfo {note} {[Erratum:
  Astron.Astrophys. 652, C4 (2021)]},\ \Eprint
  {http://arxiv.org/abs/1807.06209} {arXiv:1807.06209 [astro-ph.CO]}
  \BibitemShut {NoStop}%
\bibitem [{\citenamefont {Battaglieri}\ \emph {et~al.}(2017)\citenamefont
  {Battaglieri} \emph {et~al.}}]{Battaglieri:2017aum}%
  \BibitemOpen
  \bibfield  {author} {\bibinfo {author} {\bibfnamefont {M.}~\bibnamefont
  {Battaglieri}} \emph {et~al.},\ }in\ \href
  {http://lss.fnal.gov/archive/2017/conf/fermilab-conf-17-282-ae-ppd-t.pdf}
  {\emph {\bibinfo {booktitle} {{U.S. Cosmic Visions: New Ideas in Dark Matter
  College Park, MD, USA, March 23-25, 2017}}}}\ (\bibinfo {year} {2017})\
  \Eprint {http://arxiv.org/abs/1707.04591} {arXiv:1707.04591 [hep-ph]}
  \BibitemShut {NoStop}%
%%CITATION = ARXIV:1707.04591;%%
\bibitem [{\citenamefont {Jungman}\ \emph {et~al.}(1996)\citenamefont
  {Jungman}, \citenamefont {Kamionkowski},\ and\ \citenamefont
  {Griest}}]{Jungman:1995df}%
  \BibitemOpen
  \bibfield  {author} {\bibinfo {author} {\bibfnamefont {G.}~\bibnamefont
  {Jungman}}, \bibinfo {author} {\bibfnamefont {M.}~\bibnamefont
  {Kamionkowski}}, \ and\ \bibinfo {author} {\bibfnamefont {K.}~\bibnamefont
  {Griest}},\ }\href {\doibase 10.1016/0370-1573(95)00058-5} {\bibfield
  {journal} {\bibinfo  {journal} {Phys. Rept.}\ }\textbf {\bibinfo {volume}
  {267}},\ \bibinfo {pages} {195} (\bibinfo {year} {1996})},\ \Eprint
  {http://arxiv.org/abs/hep-ph/9506380} {arXiv:hep-ph/9506380 [hep-ph]}
  \BibitemShut {NoStop}%
%%CITATION = HEP-PH/9506380;%%
\bibitem [{\citenamefont {Bertone}\ \emph {et~al.}(2005)\citenamefont
  {Bertone}, \citenamefont {Hooper},\ and\ \citenamefont
  {Silk}}]{Bertone:2004pz}%
  \BibitemOpen
  \bibfield  {author} {\bibinfo {author} {\bibfnamefont {G.}~\bibnamefont
  {Bertone}}, \bibinfo {author} {\bibfnamefont {D.}~\bibnamefont {Hooper}}, \
  and\ \bibinfo {author} {\bibfnamefont {J.}~\bibnamefont {Silk}},\ }\href
  {\doibase 10.1016/j.physrep.2004.08.031} {\bibfield  {journal} {\bibinfo
  {journal} {Phys. Rept.}\ }\textbf {\bibinfo {volume} {405}},\ \bibinfo
  {pages} {279} (\bibinfo {year} {2005})},\ \Eprint
  {http://arxiv.org/abs/hep-ph/0404175} {arXiv:hep-ph/0404175 [hep-ph]}
  \BibitemShut {NoStop}%
%%CITATION = HEP-PH/0404175;%%
\bibitem [{\citenamefont {Feng}(2010)}]{Feng:2010gw}%
  \BibitemOpen
  \bibfield  {author} {\bibinfo {author} {\bibfnamefont {J.~L.}\ \bibnamefont
  {Feng}},\ }\href {\doibase 10.1146/annurev-astro-082708-101659} {\bibfield
  {journal} {\bibinfo  {journal} {Ann. Rev. Astron. Astrophys.}\ }\textbf
  {\bibinfo {volume} {48}},\ \bibinfo {pages} {495} (\bibinfo {year} {2010})},\
  \Eprint {http://arxiv.org/abs/1003.0904} {arXiv:1003.0904 [astro-ph.CO]}
  \BibitemShut {NoStop}%
%%CITATION = ARXIV:1003.0904;%%
\bibitem [{\citenamefont {Roszkowski}\ \emph {et~al.}(2018)\citenamefont
  {Roszkowski}, \citenamefont {Sessolo},\ and\ \citenamefont
  {Trojanowski}}]{Roszkowski:2017nbc}%
  \BibitemOpen
  \bibfield  {author} {\bibinfo {author} {\bibfnamefont {L.}~\bibnamefont
  {Roszkowski}}, \bibinfo {author} {\bibfnamefont {E.~M.}\ \bibnamefont
  {Sessolo}}, \ and\ \bibinfo {author} {\bibfnamefont {S.}~\bibnamefont
  {Trojanowski}},\ }\href {\doibase 10.1088/1361-6633/aab913} {\bibfield
  {journal} {\bibinfo  {journal} {Rept. Prog. Phys.}\ }\textbf {\bibinfo
  {volume} {81}},\ \bibinfo {pages} {066201} (\bibinfo {year} {2018})},\
  \Eprint {http://arxiv.org/abs/1707.06277} {arXiv:1707.06277 [hep-ph]}
  \BibitemShut {NoStop}%
%%CITATION = ARXIV:1707.06277;%%
\bibitem [{\citenamefont {Lin}(2019)}]{Lin:2019uvt}%
  \BibitemOpen
  \bibfield  {author} {\bibinfo {author} {\bibfnamefont {T.}~\bibnamefont
  {Lin}},\ }\href {\doibase 10.22323/1.333.0009} {\bibfield  {journal}
  {\bibinfo  {journal} {PoS}\ }\textbf {\bibinfo {volume} {333}},\ \bibinfo
  {pages} {009} (\bibinfo {year} {2019})},\ \Eprint
  {http://arxiv.org/abs/1904.07915} {arXiv:1904.07915 [hep-ph]} \BibitemShut
  {NoStop}%
\bibitem [{\citenamefont {Cooley}\ \emph {et~al.}(2022)\citenamefont {Cooley}
  \emph {et~al.}}]{Cooley:2022ufh}%
  \BibitemOpen
  \bibfield  {author} {\bibinfo {author} {\bibfnamefont {J.}~\bibnamefont
  {Cooley}} \emph {et~al.},\ }\href@noop {} {\  (\bibinfo {year} {2022})},\
  \Eprint {http://arxiv.org/abs/2209.07426} {arXiv:2209.07426 [hep-ph]}
  \BibitemShut {NoStop}%
\bibitem [{\citenamefont {Lee}\ and\ \citenamefont
  {Weinberg}(1977)}]{Lee:1977ua}%
  \BibitemOpen
  \bibfield  {author} {\bibinfo {author} {\bibfnamefont {B.~W.}\ \bibnamefont
  {Lee}}\ and\ \bibinfo {author} {\bibfnamefont {S.}~\bibnamefont {Weinberg}},\
  }\href {\doibase 10.1103/PhysRevLett.39.165} {\bibfield  {journal} {\bibinfo
  {journal} {Phys. Rev. Lett.}\ }\textbf {\bibinfo {volume} {39}},\ \bibinfo
  {pages} {165} (\bibinfo {year} {1977})}\BibitemShut {NoStop}%
%%CITATION = PRLTA,39,165;%%
\bibitem [{\citenamefont {Goodman}\ and\ \citenamefont
  {Witten}(1985)}]{Goodman:1984dc}%
  \BibitemOpen
  \bibfield  {author} {\bibinfo {author} {\bibfnamefont {M.~W.}\ \bibnamefont
  {Goodman}}\ and\ \bibinfo {author} {\bibfnamefont {E.}~\bibnamefont
  {Witten}},\ }\href {\doibase 10.1103/PhysRevD.31.3059} {\bibfield  {journal}
  {\bibinfo  {journal} {Phys. Rev.}\ }\textbf {\bibinfo {volume} {D31}},\
  \bibinfo {pages} {3059} (\bibinfo {year} {1985})}\BibitemShut {NoStop}%
%%CITATION = PHRVA,D31,3059;%%
\bibitem [{\citenamefont {Drukier}\ \emph {et~al.}(1986)\citenamefont
  {Drukier}, \citenamefont {Freese},\ and\ \citenamefont
  {Spergel}}]{Drukier:1986tm}%
  \BibitemOpen
  \bibfield  {author} {\bibinfo {author} {\bibfnamefont {A.~K.}\ \bibnamefont
  {Drukier}}, \bibinfo {author} {\bibfnamefont {K.}~\bibnamefont {Freese}}, \
  and\ \bibinfo {author} {\bibfnamefont {D.~N.}\ \bibnamefont {Spergel}},\
  }\href {\doibase 10.1103/PhysRevD.33.3495} {\bibfield  {journal} {\bibinfo
  {journal} {Phys. Rev.}\ }\textbf {\bibinfo {volume} {D33}},\ \bibinfo {pages}
  {3495} (\bibinfo {year} {1986})}\BibitemShut {NoStop}%
%%CITATION = PHRVA,D33,3495;%%
\bibitem [{\citenamefont {Billard}\ \emph {et~al.}(2022)\citenamefont {Billard}
  \emph {et~al.}}]{Billard:2021uyg}%
  \BibitemOpen
  \bibfield  {author} {\bibinfo {author} {\bibfnamefont {J.}~\bibnamefont
  {Billard}} \emph {et~al.},\ }\href {\doibase 10.1088/1361-6633/ac5754}
  {\bibfield  {journal} {\bibinfo  {journal} {Rept. Prog. Phys.}\ }\textbf
  {\bibinfo {volume} {85}},\ \bibinfo {pages} {056201} (\bibinfo {year}
  {2022})},\ \Eprint {http://arxiv.org/abs/2104.07634} {arXiv:2104.07634
  [hep-ex]} \BibitemShut {NoStop}%
\bibitem [{\citenamefont {Meng}\ \emph {et~al.}(2021)\citenamefont {Meng} \emph
  {et~al.}}]{PandaX-4T:2021bab}%
  \BibitemOpen
  \bibfield  {author} {\bibinfo {author} {\bibfnamefont {Y.}~\bibnamefont
  {Meng}} \emph {et~al.} (\bibinfo {collaboration} {PandaX-4T}),\ }\href
  {\doibase 10.1103/PhysRevLett.127.261802} {\bibfield  {journal} {\bibinfo
  {journal} {Phys. Rev. Lett.}\ }\textbf {\bibinfo {volume} {127}},\ \bibinfo
  {pages} {261802} (\bibinfo {year} {2021})},\ \Eprint
  {http://arxiv.org/abs/2107.13438} {arXiv:2107.13438 [hep-ex]} \BibitemShut
  {NoStop}%
\bibitem [{\citenamefont {Aalbers}\ \emph {et~al.}(2022)\citenamefont {Aalbers}
  \emph {et~al.}}]{LZ:2022ufs}%
  \BibitemOpen
  \bibfield  {author} {\bibinfo {author} {\bibfnamefont {J.}~\bibnamefont
  {Aalbers}} \emph {et~al.} (\bibinfo {collaboration} {LZ}),\ }\href@noop {} {\
   (\bibinfo {year} {2022})},\ \Eprint {http://arxiv.org/abs/2207.03764}
  {arXiv:2207.03764 [hep-ex]} \BibitemShut {NoStop}%
\bibitem [{\citenamefont {Aprile}\ \emph {et~al.}(2023)\citenamefont {Aprile}
  \emph {et~al.}}]{XENON:2023sxq}%
  \BibitemOpen
  \bibfield  {author} {\bibinfo {author} {\bibfnamefont {E.}~\bibnamefont
  {Aprile}} \emph {et~al.} (\bibinfo {collaboration} {XENON}),\ }\href@noop {}
  {\  (\bibinfo {year} {2023})},\ \Eprint {http://arxiv.org/abs/2303.14729}
  {arXiv:2303.14729 [hep-ex]} \BibitemShut {NoStop}%
\bibitem [{\citenamefont {Ajaj}\ \emph {et~al.}(2019)\citenamefont {Ajaj} \emph
  {et~al.}}]{DEAP:2019yzn}%
  \BibitemOpen
  \bibfield  {author} {\bibinfo {author} {\bibfnamefont {R.}~\bibnamefont
  {Ajaj}} \emph {et~al.} (\bibinfo {collaboration} {DEAP}),\ }\href {\doibase
  10.1103/PhysRevD.100.022004} {\bibfield  {journal} {\bibinfo  {journal}
  {Phys. Rev. D}\ }\textbf {\bibinfo {volume} {100}},\ \bibinfo {pages}
  {022004} (\bibinfo {year} {2019})},\ \Eprint
  {http://arxiv.org/abs/1902.04048} {arXiv:1902.04048 [astro-ph.CO]}
  \BibitemShut {NoStop}%
\bibitem [{\citenamefont {Aalseth}\ \emph {et~al.}(2018)\citenamefont {Aalseth}
  \emph {et~al.}}]{Aalseth:2017fik}%
  \BibitemOpen
  \bibfield  {author} {\bibinfo {author} {\bibfnamefont {C.~E.}\ \bibnamefont
  {Aalseth}} \emph {et~al.},\ }\href {\doibase 10.1140/epjp/i2018-11973-4}
  {\bibfield  {journal} {\bibinfo  {journal} {Eur. Phys. J. Plus}\ }\textbf
  {\bibinfo {volume} {133}},\ \bibinfo {pages} {131} (\bibinfo {year}
  {2018})},\ \Eprint {http://arxiv.org/abs/1707.08145} {arXiv:1707.08145
  [physics.ins-det]} \BibitemShut {NoStop}%
%%CITATION = ARXIV:1707.08145;%%
\bibitem [{\citenamefont {Freedman}(1974)}]{Freedman:1973yd}%
  \BibitemOpen
  \bibfield  {author} {\bibinfo {author} {\bibfnamefont {D.~Z.}\ \bibnamefont
  {Freedman}},\ }\href {\doibase 10.1103/PhysRevD.9.1389} {\bibfield  {journal}
  {\bibinfo  {journal} {Phys. Rev.}\ }\textbf {\bibinfo {volume} {D9}},\
  \bibinfo {pages} {1389} (\bibinfo {year} {1974})}\BibitemShut {NoStop}%
%%CITATION = PHRVA,D9,1389;%%
\bibitem [{\citenamefont {Monroe}\ and\ \citenamefont
  {Fisher}(2007)}]{Monroe:2007xp}%
  \BibitemOpen
  \bibfield  {author} {\bibinfo {author} {\bibfnamefont {J.}~\bibnamefont
  {Monroe}}\ and\ \bibinfo {author} {\bibfnamefont {P.}~\bibnamefont
  {Fisher}},\ }\href {\doibase 10.1103/PhysRevD.76.033007} {\bibfield
  {journal} {\bibinfo  {journal} {Phys. Rev.}\ }\textbf {\bibinfo {volume}
  {D76}},\ \bibinfo {pages} {033007} (\bibinfo {year} {2007})},\ \Eprint
  {http://arxiv.org/abs/0706.3019} {arXiv:0706.3019 [astro-ph]} \BibitemShut
  {NoStop}%
%%CITATION = ARXIV:0706.3019;%%
\bibitem [{\citenamefont {Dodelson}(2009)}]{Dodelson:2008yx}%
  \BibitemOpen
  \bibfield  {author} {\bibinfo {author} {\bibfnamefont {S.}~\bibnamefont
  {Dodelson}},\ }\href {\doibase 10.1103/PhysRevD.79.043508} {\bibfield
  {journal} {\bibinfo  {journal} {Phys. Rev.}\ }\textbf {\bibinfo {volume}
  {D79}},\ \bibinfo {pages} {043508} (\bibinfo {year} {2009})},\ \Eprint
  {http://arxiv.org/abs/0812.0787} {arXiv:0812.0787 [astro-ph]} \BibitemShut
  {NoStop}%
%%CITATION = ARXIV:0812.0787;%%
\bibitem [{\citenamefont {Strigari}(2009)}]{Strigari:2009bq}%
  \BibitemOpen
  \bibfield  {author} {\bibinfo {author} {\bibfnamefont {L.~E.}\ \bibnamefont
  {Strigari}},\ }\href {\doibase 10.1088/1367-2630/11/10/105011} {\bibfield
  {journal} {\bibinfo  {journal} {New J. Phys.}\ }\textbf {\bibinfo {volume}
  {11}},\ \bibinfo {pages} {105011} (\bibinfo {year} {2009})},\ \Eprint
  {http://arxiv.org/abs/0903.3630} {arXiv:0903.3630 [astro-ph.CO]} \BibitemShut
  {NoStop}%
%%CITATION = ARXIV:0903.3630;%%
\bibitem [{\citenamefont {Billard}\ \emph {et~al.}(2014)\citenamefont
  {Billard}, \citenamefont {Strigari},\ and\ \citenamefont
  {Figueroa-Feliciano}}]{Billard:2013qya}%
  \BibitemOpen
  \bibfield  {author} {\bibinfo {author} {\bibfnamefont {J.}~\bibnamefont
  {Billard}}, \bibinfo {author} {\bibfnamefont {L.}~\bibnamefont {Strigari}}, \
  and\ \bibinfo {author} {\bibfnamefont {E.}~\bibnamefont
  {Figueroa-Feliciano}},\ }\href {\doibase 10.1103/PhysRevD.89.023524}
  {\bibfield  {journal} {\bibinfo  {journal} {Phys. Rev.}\ }\textbf {\bibinfo
  {volume} {D89}},\ \bibinfo {pages} {023524} (\bibinfo {year} {2014})},\
  \Eprint {http://arxiv.org/abs/1307.5458} {arXiv:1307.5458 [hep-ph]}
  \BibitemShut {NoStop}%
%%CITATION = ARXIV:1307.5458;%%
\bibitem [{\citenamefont {Ruppin}\ \emph {et~al.}(2014)\citenamefont {Ruppin},
  \citenamefont {Billard}, \citenamefont {Figueroa-Feliciano},\ and\
  \citenamefont {Strigari}}]{Ruppin:2014bra}%
  \BibitemOpen
  \bibfield  {author} {\bibinfo {author} {\bibfnamefont {F.}~\bibnamefont
  {Ruppin}}, \bibinfo {author} {\bibfnamefont {J.}~\bibnamefont {Billard}},
  \bibinfo {author} {\bibfnamefont {E.}~\bibnamefont {Figueroa-Feliciano}}, \
  and\ \bibinfo {author} {\bibfnamefont {L.}~\bibnamefont {Strigari}},\ }\href
  {\doibase 10.1103/PhysRevD.90.083510} {\bibfield  {journal} {\bibinfo
  {journal} {Phys. Rev.}\ }\textbf {\bibinfo {volume} {D90}},\ \bibinfo {pages}
  {083510} (\bibinfo {year} {2014})},\ \Eprint {http://arxiv.org/abs/1408.3581}
  {arXiv:1408.3581 [hep-ph]} \BibitemShut {NoStop}%
%%CITATION = ARXIV:1408.3581;%%
\bibitem [{\citenamefont {O'Hare}(2016)}]{OHare:2016pjy}%
  \BibitemOpen
  \bibfield  {author} {\bibinfo {author} {\bibfnamefont {C.~A.~J.}\
  \bibnamefont {O'Hare}},\ }\href {\doibase 10.1103/PhysRevD.94.063527}
  {\bibfield  {journal} {\bibinfo  {journal} {Phys. Rev. D}\ }\textbf {\bibinfo
  {volume} {94}},\ \bibinfo {pages} {063527} (\bibinfo {year} {2016})},\
  \Eprint {http://arxiv.org/abs/1604.03858} {arXiv:1604.03858 [astro-ph.CO]}
  \BibitemShut {NoStop}%
\bibitem [{\citenamefont {Gelmini}\ \emph {et~al.}(2018)\citenamefont
  {Gelmini}, \citenamefont {Takhistov},\ and\ \citenamefont
  {Witte}}]{Gelmini:2018ogy}%
  \BibitemOpen
  \bibfield  {author} {\bibinfo {author} {\bibfnamefont {G.~B.}\ \bibnamefont
  {Gelmini}}, \bibinfo {author} {\bibfnamefont {V.}~\bibnamefont {Takhistov}},
  \ and\ \bibinfo {author} {\bibfnamefont {S.~J.}\ \bibnamefont {Witte}},\
  }\href {\doibase 10.1088/1475-7516/2018/07/009,
  10.1088/1475-7516/2019/02/E02} {\bibfield  {journal} {\bibinfo  {journal}
  {JCAP}\ }\textbf {\bibinfo {volume} {1807}},\ \bibinfo {pages} {009}
  (\bibinfo {year} {2018})},\ \bibinfo {note} {[Erratum: JCAP1902,E02(2019)]},\
  \Eprint {http://arxiv.org/abs/1804.01638} {arXiv:1804.01638 [hep-ph]}
  \BibitemShut {NoStop}%
%%CITATION = ARXIV:1804.01638;%%
\bibitem [{\citenamefont {O'Hare}(2020)}]{OHare:2020lva}%
  \BibitemOpen
  \bibfield  {author} {\bibinfo {author} {\bibfnamefont {C.~A.~J.}\
  \bibnamefont {O'Hare}},\ }\href {\doibase 10.1103/PhysRevD.102.063024}
  {\bibfield  {journal} {\bibinfo  {journal} {Phys. Rev.}\ }\textbf {\bibinfo
  {volume} {D102}},\ \bibinfo {pages} {063024} (\bibinfo {year} {2020})},\
  \Eprint {http://arxiv.org/abs/2002.07499} {arXiv:2002.07499 [astro-ph.CO]}
  \BibitemShut {NoStop}%
%%CITATION = ARXIV:2002.07499;%%
\bibitem [{\citenamefont {Newstead}\ \emph {et~al.}(2021)\citenamefont
  {Newstead}, \citenamefont {Lang},\ and\ \citenamefont
  {Strigari}}]{Newstead:2020fie}%
  \BibitemOpen
  \bibfield  {author} {\bibinfo {author} {\bibfnamefont {J.~L.}\ \bibnamefont
  {Newstead}}, \bibinfo {author} {\bibfnamefont {R.~F.}\ \bibnamefont {Lang}},
  \ and\ \bibinfo {author} {\bibfnamefont {L.~E.}\ \bibnamefont {Strigari}},\
  }\href {\doibase 10.1103/PhysRevD.104.115022} {\bibfield  {journal} {\bibinfo
   {journal} {Phys. Rev. D}\ }\textbf {\bibinfo {volume} {104}},\ \bibinfo
  {pages} {115022} (\bibinfo {year} {2021})},\ \Eprint
  {http://arxiv.org/abs/2002.08566} {arXiv:2002.08566 [astro-ph.CO]}
  \BibitemShut {NoStop}%
\bibitem [{\citenamefont {Gaspert}\ \emph {et~al.}(2022)\citenamefont
  {Gaspert}, \citenamefont {Giampa},\ and\ \citenamefont
  {Morrissey}}]{Gaspert:2021gyj}%
  \BibitemOpen
  \bibfield  {author} {\bibinfo {author} {\bibfnamefont {A.}~\bibnamefont
  {Gaspert}}, \bibinfo {author} {\bibfnamefont {P.}~\bibnamefont {Giampa}}, \
  and\ \bibinfo {author} {\bibfnamefont {D.~E.}\ \bibnamefont {Morrissey}},\
  }\href {\doibase 10.1103/PhysRevD.105.035020} {\bibfield  {journal} {\bibinfo
   {journal} {Phys. Rev. D}\ }\textbf {\bibinfo {volume} {105}},\ \bibinfo
  {pages} {035020} (\bibinfo {year} {2022})},\ \Eprint
  {http://arxiv.org/abs/2108.03248} {arXiv:2108.03248 [hep-ph]} \BibitemShut
  {NoStop}%
\bibitem [{\citenamefont {O'Hare}(2021)}]{OHare:2021utq}%
  \BibitemOpen
  \bibfield  {author} {\bibinfo {author} {\bibfnamefont {C.~A.~J.}\
  \bibnamefont {O'Hare}},\ }\href {\doibase 10.1103/PhysRevLett.127.251802}
  {\bibfield  {journal} {\bibinfo  {journal} {Phys. Rev. Lett.}\ }\textbf
  {\bibinfo {volume} {127}},\ \bibinfo {pages} {251802} (\bibinfo {year}
  {2021})},\ \Eprint {http://arxiv.org/abs/2109.03116} {arXiv:2109.03116
  [hep-ph]} \BibitemShut {NoStop}%
\bibitem [{\citenamefont {Akerib}\ \emph {et~al.}(2022)\citenamefont {Akerib}
  \emph {et~al.}}]{Akerib:2022ort}%
  \BibitemOpen
  \bibfield  {author} {\bibinfo {author} {\bibfnamefont {D.~S.}\ \bibnamefont
  {Akerib}} \emph {et~al.},\ }in\ \href@noop {} {\emph {\bibinfo {booktitle}
  {{2022 Snowmass Summer Study}}}}\ (\bibinfo {year} {2022})\ \Eprint
  {http://arxiv.org/abs/2203.08084} {arXiv:2203.08084 [hep-ex]} \BibitemShut
  {NoStop}%
\bibitem [{\citenamefont {Aalbers}\ \emph {et~al.}(2016)\citenamefont {Aalbers}
  \emph {et~al.}}]{DARWIN:2016hyl}%
  \BibitemOpen
  \bibfield  {author} {\bibinfo {author} {\bibfnamefont {J.}~\bibnamefont
  {Aalbers}} \emph {et~al.} (\bibinfo {collaboration} {DARWIN}),\ }\href
  {\doibase 10.1088/1475-7516/2016/11/017} {\bibfield  {journal} {\bibinfo
  {journal} {JCAP}\ }\textbf {\bibinfo {volume} {11}},\ \bibinfo {pages} {017}
  (\bibinfo {year} {2016})},\ \Eprint {http://arxiv.org/abs/1606.07001}
  {arXiv:1606.07001 [astro-ph.IM]} \BibitemShut {NoStop}%
\bibitem [{\citenamefont {Aalbers}\ \emph {et~al.}(2023)\citenamefont {Aalbers}
  \emph {et~al.}}]{Aalbers:2022dzr}%
  \BibitemOpen
  \bibfield  {author} {\bibinfo {author} {\bibfnamefont {J.}~\bibnamefont
  {Aalbers}} \emph {et~al.},\ }\href {\doibase 10.1088/1361-6471/ac841a}
  {\bibfield  {journal} {\bibinfo  {journal} {J. Phys. G}\ }\textbf {\bibinfo
  {volume} {50}},\ \bibinfo {pages} {013001} (\bibinfo {year} {2023})},\
  \Eprint {http://arxiv.org/abs/2203.02309} {arXiv:2203.02309
  [physics.ins-det]} \BibitemShut {NoStop}%
\bibitem [{\citenamefont {Galbiati}(2018)}]{Galbiati:2018}%
  \BibitemOpen
  \bibfield  {author} {\bibinfo {author} {\bibfnamefont {C.}~\bibnamefont
  {Galbiati}} (\bibinfo {collaboration} {GADMC}),\ }\href@noop {} {\enquote
  {\bibinfo {title} {\textit{Future Dark Matter Searches with Low-Radioactivity
  Argon}},}\ }\bibinfo {howpublished}
  {\url{https://indico.cern.ch/event/765096/contributions/3295671/}} (\bibinfo
  {year} {2018})\BibitemShut {NoStop}%
\bibitem [{\citenamefont {Ellis}\ \emph {et~al.}(2005)\citenamefont {Ellis},
  \citenamefont {Olive}, \citenamefont {Santoso},\ and\ \citenamefont
  {Spanos}}]{Ellis:2005mb}%
  \BibitemOpen
  \bibfield  {author} {\bibinfo {author} {\bibfnamefont {J.~R.}\ \bibnamefont
  {Ellis}}, \bibinfo {author} {\bibfnamefont {K.~A.}\ \bibnamefont {Olive}},
  \bibinfo {author} {\bibfnamefont {Y.}~\bibnamefont {Santoso}}, \ and\
  \bibinfo {author} {\bibfnamefont {V.~C.}\ \bibnamefont {Spanos}},\ }\href
  {\doibase 10.1103/PhysRevD.71.095007} {\bibfield  {journal} {\bibinfo
  {journal} {Phys. Rev.}\ }\textbf {\bibinfo {volume} {D71}},\ \bibinfo {pages}
  {095007} (\bibinfo {year} {2005})},\ \Eprint
  {http://arxiv.org/abs/hep-ph/0502001} {arXiv:hep-ph/0502001 [hep-ph]}
  \BibitemShut {NoStop}%
%%CITATION = HEP-PH/0502001;%%
\bibitem [{\citenamefont {Cheung}\ \emph {et~al.}(2013)\citenamefont {Cheung},
  \citenamefont {Hall}, \citenamefont {Pinner},\ and\ \citenamefont
  {Ruderman}}]{Cheung:2012qy}%
  \BibitemOpen
  \bibfield  {author} {\bibinfo {author} {\bibfnamefont {C.}~\bibnamefont
  {Cheung}}, \bibinfo {author} {\bibfnamefont {L.~J.}\ \bibnamefont {Hall}},
  \bibinfo {author} {\bibfnamefont {D.}~\bibnamefont {Pinner}}, \ and\ \bibinfo
  {author} {\bibfnamefont {J.~T.}\ \bibnamefont {Ruderman}},\ }\href {\doibase
  10.1007/JHEP05(2013)100} {\bibfield  {journal} {\bibinfo  {journal} {JHEP}\
  }\textbf {\bibinfo {volume} {05}},\ \bibinfo {pages} {100} (\bibinfo {year}
  {2013})},\ \Eprint {http://arxiv.org/abs/1211.4873} {arXiv:1211.4873
  [hep-ph]} \BibitemShut {NoStop}%
%%CITATION = ARXIV:1211.4873;%%
\bibitem [{\citenamefont {Hill}\ and\ \citenamefont
  {Solon}(2014)}]{Hill:2013hoa}%
  \BibitemOpen
  \bibfield  {author} {\bibinfo {author} {\bibfnamefont {R.~J.}\ \bibnamefont
  {Hill}}\ and\ \bibinfo {author} {\bibfnamefont {M.~P.}\ \bibnamefont
  {Solon}},\ }\href {\doibase 10.1103/PhysRevLett.112.211602} {\bibfield
  {journal} {\bibinfo  {journal} {Phys. Rev. Lett.}\ }\textbf {\bibinfo
  {volume} {112}},\ \bibinfo {pages} {211602} (\bibinfo {year} {2014})},\
  \Eprint {http://arxiv.org/abs/1309.4092} {arXiv:1309.4092 [hep-ph]}
  \BibitemShut {NoStop}%
%%CITATION = ARXIV:1309.4092;%%
\bibitem [{\citenamefont {Cahill-Rowley}\ \emph {et~al.}(2015)\citenamefont
  {Cahill-Rowley}, \citenamefont {Cotta}, \citenamefont {Drlica-Wagner},
  \citenamefont {Funk}, \citenamefont {Hewett}, \citenamefont {Ismail},
  \citenamefont {Rizzo},\ and\ \citenamefont {Wood}}]{Cahill-Rowley:2014boa}%
  \BibitemOpen
  \bibfield  {author} {\bibinfo {author} {\bibfnamefont {M.}~\bibnamefont
  {Cahill-Rowley}}, \bibinfo {author} {\bibfnamefont {R.}~\bibnamefont
  {Cotta}}, \bibinfo {author} {\bibfnamefont {A.}~\bibnamefont
  {Drlica-Wagner}}, \bibinfo {author} {\bibfnamefont {S.}~\bibnamefont {Funk}},
  \bibinfo {author} {\bibfnamefont {J.}~\bibnamefont {Hewett}}, \bibinfo
  {author} {\bibfnamefont {A.}~\bibnamefont {Ismail}}, \bibinfo {author}
  {\bibfnamefont {T.}~\bibnamefont {Rizzo}}, \ and\ \bibinfo {author}
  {\bibfnamefont {M.}~\bibnamefont {Wood}},\ }\href {\doibase
  10.1103/PhysRevD.91.055011} {\bibfield  {journal} {\bibinfo  {journal} {Phys.
  Rev.}\ }\textbf {\bibinfo {volume} {D91}},\ \bibinfo {pages} {055011}
  (\bibinfo {year} {2015})},\ \Eprint {http://arxiv.org/abs/1405.6716}
  {arXiv:1405.6716 [hep-ph]} \BibitemShut {NoStop}%
%%CITATION = ARXIV:1405.6716;%%
\bibitem [{\citenamefont {Bramante}\ \emph {et~al.}(2016)\citenamefont
  {Bramante}, \citenamefont {Desai}, \citenamefont {Fox}, \citenamefont
  {Martin}, \citenamefont {Ostdiek},\ and\ \citenamefont
  {Plehn}}]{Bramante:2015una}%
  \BibitemOpen
  \bibfield  {author} {\bibinfo {author} {\bibfnamefont {J.}~\bibnamefont
  {Bramante}}, \bibinfo {author} {\bibfnamefont {N.}~\bibnamefont {Desai}},
  \bibinfo {author} {\bibfnamefont {P.}~\bibnamefont {Fox}}, \bibinfo {author}
  {\bibfnamefont {A.}~\bibnamefont {Martin}}, \bibinfo {author} {\bibfnamefont
  {B.}~\bibnamefont {Ostdiek}}, \ and\ \bibinfo {author} {\bibfnamefont
  {T.}~\bibnamefont {Plehn}},\ }\href {\doibase 10.1103/PhysRevD.93.063525}
  {\bibfield  {journal} {\bibinfo  {journal} {Phys. Rev.}\ }\textbf {\bibinfo
  {volume} {D93}},\ \bibinfo {pages} {063525} (\bibinfo {year} {2016})},\
  \Eprint {http://arxiv.org/abs/1510.03460} {arXiv:1510.03460 [hep-ph]}
  \BibitemShut {NoStop}%
%%CITATION = ARXIV:1510.03460;%%
\bibitem [{\citenamefont {Claude}\ and\ \citenamefont
  {Godfrey}(2021)}]{Claude:2021sye}%
  \BibitemOpen
  \bibfield  {author} {\bibinfo {author} {\bibfnamefont {J.}~\bibnamefont
  {Claude}}\ and\ \bibinfo {author} {\bibfnamefont {S.}~\bibnamefont
  {Godfrey}},\ }\href {\doibase 10.1140/epjc/s10052-021-09170-0} {\bibfield
  {journal} {\bibinfo  {journal} {Eur. Phys. J.}\ }\textbf {\bibinfo {volume}
  {C81}},\ \bibinfo {pages} {405} (\bibinfo {year} {2021})},\ \Eprint
  {http://arxiv.org/abs/2104.01096} {arXiv:2104.01096 [hep-ph]} \BibitemShut
  {NoStop}%
%%CITATION = ARXIV:2104.01096;%%
\bibitem [{\citenamefont {Hill}\ and\ \citenamefont
  {Solon}(2015)}]{Hill:2014yxa}%
  \BibitemOpen
  \bibfield  {author} {\bibinfo {author} {\bibfnamefont {R.~J.}\ \bibnamefont
  {Hill}}\ and\ \bibinfo {author} {\bibfnamefont {M.~P.}\ \bibnamefont
  {Solon}},\ }\href {\doibase 10.1103/PhysRevD.91.043505} {\bibfield  {journal}
  {\bibinfo  {journal} {Phys. Rev. D}\ }\textbf {\bibinfo {volume} {91}},\
  \bibinfo {pages} {043505} (\bibinfo {year} {2015})},\ \Eprint
  {http://arxiv.org/abs/1409.8290} {arXiv:1409.8290 [hep-ph]} \BibitemShut
  {NoStop}%
\bibitem [{\citenamefont {Co}\ \emph {et~al.}(2022)\citenamefont {Co},
  \citenamefont {Sheff},\ and\ \citenamefont {Wells}}]{Co:2021ion}%
  \BibitemOpen
  \bibfield  {author} {\bibinfo {author} {\bibfnamefont {R.~T.}\ \bibnamefont
  {Co}}, \bibinfo {author} {\bibfnamefont {B.}~\bibnamefont {Sheff}}, \ and\
  \bibinfo {author} {\bibfnamefont {J.~D.}\ \bibnamefont {Wells}},\ }\href
  {\doibase 10.1103/PhysRevD.105.035012} {\bibfield  {journal} {\bibinfo
  {journal} {Phys. Rev. D}\ }\textbf {\bibinfo {volume} {105}},\ \bibinfo
  {pages} {035012} (\bibinfo {year} {2022})},\ \Eprint
  {http://arxiv.org/abs/2105.12142} {arXiv:2105.12142 [hep-ph]} \BibitemShut
  {NoStop}%
\bibitem [{\citenamefont {Dessert}\ \emph {et~al.}(2022)\citenamefont
  {Dessert}, \citenamefont {Foster}, \citenamefont {Park}, \citenamefont
  {Safdi},\ and\ \citenamefont {Xu}}]{Dessert:2022evk}%
  \BibitemOpen
  \bibfield  {author} {\bibinfo {author} {\bibfnamefont {C.}~\bibnamefont
  {Dessert}}, \bibinfo {author} {\bibfnamefont {J.~W.}\ \bibnamefont {Foster}},
  \bibinfo {author} {\bibfnamefont {Y.}~\bibnamefont {Park}}, \bibinfo {author}
  {\bibfnamefont {B.~R.}\ \bibnamefont {Safdi}}, \ and\ \bibinfo {author}
  {\bibfnamefont {W.~L.}\ \bibnamefont {Xu}},\ }\href@noop {} {\  (\bibinfo
  {year} {2022})},\ \Eprint {http://arxiv.org/abs/2207.10090} {arXiv:2207.10090
  [hep-ph]} \BibitemShut {NoStop}%
\bibitem [{\citenamefont {O'Hare}\ \emph {et~al.}(2015)\citenamefont {O'Hare},
  \citenamefont {Green}, \citenamefont {Billard}, \citenamefont
  {Figueroa-Feliciano},\ and\ \citenamefont {Strigari}}]{OHare:2015utx}%
  \BibitemOpen
  \bibfield  {author} {\bibinfo {author} {\bibfnamefont {C.~A.~J.}\
  \bibnamefont {O'Hare}}, \bibinfo {author} {\bibfnamefont {A.~M.}\
  \bibnamefont {Green}}, \bibinfo {author} {\bibfnamefont {J.}~\bibnamefont
  {Billard}}, \bibinfo {author} {\bibfnamefont {E.}~\bibnamefont
  {Figueroa-Feliciano}}, \ and\ \bibinfo {author} {\bibfnamefont {L.~E.}\
  \bibnamefont {Strigari}},\ }\href {\doibase 10.1103/PhysRevD.92.063518}
  {\bibfield  {journal} {\bibinfo  {journal} {Phys. Rev. D}\ }\textbf {\bibinfo
  {volume} {92}},\ \bibinfo {pages} {063518} (\bibinfo {year} {2015})},\
  \Eprint {http://arxiv.org/abs/1505.08061} {arXiv:1505.08061 [astro-ph.CO]}
  \BibitemShut {NoStop}%
\bibitem [{\citenamefont {Mayet}\ \emph {et~al.}(2016)\citenamefont {Mayet}
  \emph {et~al.}}]{Mayet:2016zxu}%
  \BibitemOpen
  \bibfield  {author} {\bibinfo {author} {\bibfnamefont {F.}~\bibnamefont
  {Mayet}} \emph {et~al.},\ }\href {\doibase 10.1016/j.physrep.2016.02.007}
  {\bibfield  {journal} {\bibinfo  {journal} {Phys. Rept.}\ }\textbf {\bibinfo
  {volume} {627}},\ \bibinfo {pages} {1} (\bibinfo {year} {2016})},\ \Eprint
  {http://arxiv.org/abs/1602.03781} {arXiv:1602.03781 [astro-ph.CO]}
  \BibitemShut {NoStop}%
\bibitem [{\citenamefont {O'Hare}\ \emph {et~al.}(2022)\citenamefont {O'Hare}
  \emph {et~al.}}]{OHare:2022jnx}%
  \BibitemOpen
  \bibfield  {author} {\bibinfo {author} {\bibfnamefont {C.~A.~J.}\
  \bibnamefont {O'Hare}} \emph {et~al.},\ }in\ \href@noop {} {\emph {\bibinfo
  {booktitle} {{Snowmass 2021}}}}\ (\bibinfo {year} {2022})\ \Eprint
  {http://arxiv.org/abs/2203.05914} {arXiv:2203.05914 [physics.ins-det]}
  \BibitemShut {NoStop}%
\bibitem [{\citenamefont {Barr}\ \emph {et~al.}(1989)\citenamefont {Barr},
  \citenamefont {Gaisser},\ and\ \citenamefont {Stanev}}]{Barr:1989ru}%
  \BibitemOpen
  \bibfield  {author} {\bibinfo {author} {\bibfnamefont {G.}~\bibnamefont
  {Barr}}, \bibinfo {author} {\bibfnamefont {T.~K.}\ \bibnamefont {Gaisser}}, \
  and\ \bibinfo {author} {\bibfnamefont {T.}~\bibnamefont {Stanev}},\ }\href
  {\doibase 10.1103/PhysRevD.39.3532} {\bibfield  {journal} {\bibinfo
  {journal} {Phys. Rev.}\ }\textbf {\bibinfo {volume} {D39}},\ \bibinfo {pages}
  {3532} (\bibinfo {year} {1989})}\BibitemShut {NoStop}%
%%CITATION = PHRVA,D39,3532;%%
\bibitem [{\citenamefont {{Volkova}}\ and\ \citenamefont
  {{Zatsepin}}(1965)}]{Volkova:1965}%
  \BibitemOpen
  \bibfield  {author} {\bibinfo {author} {\bibfnamefont {L.~V.}\ \bibnamefont
  {{Volkova}}}\ and\ \bibinfo {author} {\bibfnamefont {G.~T.}\ \bibnamefont
  {{Zatsepin}}},\ }in\ \href@noop {} {\emph {\bibinfo {booktitle}
  {International Cosmic Ray Conference}}},\ \bibinfo {series} {International
  Cosmic Ray Conference}, Vol.~\bibinfo {volume} {1}\ (\bibinfo {year} {1965})\
  p.\ \bibinfo {pages} {1039}\BibitemShut {NoStop}%
\bibitem [{\citenamefont {Miller}\ and\ \citenamefont
  {Cohen}(2006)}]{Miller:2006}%
  \BibitemOpen
  \bibfield  {author} {\bibinfo {author} {\bibfnamefont {R.}~\bibnamefont
  {Miller}}\ and\ \bibinfo {author} {\bibfnamefont {T.}~\bibnamefont {Cohen}},\
  }\href {\doibase 10.1016/j.astropartphys.2006.03.009} {\bibfield  {journal}
  {\bibinfo  {journal} {Astroparticle Physics}\ }\textbf {\bibinfo {volume}
  {25}},\ \bibinfo {pages} {368} (\bibinfo {year} {2006})}\BibitemShut
  {NoStop}%
\bibitem [{\citenamefont {Demidov}\ and\ \citenamefont
  {Gorbunov}(2021)}]{Demidov:2020bff}%
  \BibitemOpen
  \bibfield  {author} {\bibinfo {author} {\bibfnamefont {S.}~\bibnamefont
  {Demidov}}\ and\ \bibinfo {author} {\bibfnamefont {D.}~\bibnamefont
  {Gorbunov}},\ }\href {\doibase 10.1103/PhysRevD.104.043023} {\bibfield
  {journal} {\bibinfo  {journal} {Phys. Rev. D}\ }\textbf {\bibinfo {volume}
  {104}},\ \bibinfo {pages} {043023} (\bibinfo {year} {2021})},\ \Eprint
  {http://arxiv.org/abs/2012.12870} {arXiv:2012.12870 [hep-ph]} \BibitemShut
  {NoStop}%
\bibitem [{\citenamefont {Bahcall}(2005)}]{bahcallshape}%
  \BibitemOpen
  \bibfield  {author} {\bibinfo {author} {\bibfnamefont {J.}~\bibnamefont
  {Bahcall}},\ }\href {http://www.sns.ias.edu/~jnb/SNdata/sndata.html}
  {\enquote {\bibinfo {title} {{Software and data for solar neutrino
  research}},}\ } (\bibinfo {year} {2005})\BibitemShut {NoStop}%
\bibitem [{\citenamefont {Haxton}\ \emph {et~al.}(2013)\citenamefont {Haxton},
  \citenamefont {Hamish~Robertson},\ and\ \citenamefont
  {Serenelli}}]{Robertson:2012ib}%
  \BibitemOpen
  \bibfield  {author} {\bibinfo {author} {\bibfnamefont {W.~C.}\ \bibnamefont
  {Haxton}}, \bibinfo {author} {\bibfnamefont {R.~G.}\ \bibnamefont
  {Hamish~Robertson}}, \ and\ \bibinfo {author} {\bibfnamefont {A.~M.}\
  \bibnamefont {Serenelli}},\ }\href {\doibase
  10.1146/annurev-astro-081811-125539} {\bibfield  {journal} {\bibinfo
  {journal} {Ann. Rev. Astron. Astrophys.}\ }\textbf {\bibinfo {volume} {51}},\
  \bibinfo {pages} {21} (\bibinfo {year} {2013})},\ \Eprint
  {http://arxiv.org/abs/1208.5723} {arXiv:1208.5723 [astro-ph.SR]} \BibitemShut
  {NoStop}%
%%CITATION = ARXIV:1208.5723;%%
\bibitem [{\citenamefont {Vinyoles}\ \emph {et~al.}(2017)\citenamefont
  {Vinyoles}, \citenamefont {Serenelli}, \citenamefont {Villante},
  \citenamefont {Basu}, \citenamefont {Bergstr\"om}, \citenamefont
  {Gonzalez-Garcia}, \citenamefont {Maltoni}, \citenamefont {Pe\~na Garay},\
  and\ \citenamefont {Song}}]{Vinyoles:2016djt}%
  \BibitemOpen
  \bibfield  {author} {\bibinfo {author} {\bibfnamefont {N.}~\bibnamefont
  {Vinyoles}}, \bibinfo {author} {\bibfnamefont {A.~M.}\ \bibnamefont
  {Serenelli}}, \bibinfo {author} {\bibfnamefont {F.~L.}\ \bibnamefont
  {Villante}}, \bibinfo {author} {\bibfnamefont {S.}~\bibnamefont {Basu}},
  \bibinfo {author} {\bibfnamefont {J.}~\bibnamefont {Bergstr\"om}}, \bibinfo
  {author} {\bibfnamefont {M.~C.}\ \bibnamefont {Gonzalez-Garcia}}, \bibinfo
  {author} {\bibfnamefont {M.}~\bibnamefont {Maltoni}}, \bibinfo {author}
  {\bibfnamefont {C.}~\bibnamefont {Pe\~na Garay}}, \ and\ \bibinfo {author}
  {\bibfnamefont {N.}~\bibnamefont {Song}},\ }\href {\doibase
  10.3847/1538-4357/835/2/202} {\bibfield  {journal} {\bibinfo  {journal}
  {Astrophys. J.}\ }\textbf {\bibinfo {volume} {835}},\ \bibinfo {pages} {202}
  (\bibinfo {year} {2017})},\ \Eprint {http://arxiv.org/abs/1611.09867}
  {arXiv:1611.09867 [astro-ph.SR]} \BibitemShut {NoStop}%
\bibitem [{\citenamefont {Beacom}(2010)}]{Beacom:2010kk}%
  \BibitemOpen
  \bibfield  {author} {\bibinfo {author} {\bibfnamefont {J.~F.}\ \bibnamefont
  {Beacom}},\ }\href {\doibase 10.1146/annurev.nucl.010909.083331} {\bibfield
  {journal} {\bibinfo  {journal} {Ann. Rev. Nucl. Part. Sci.}\ }\textbf
  {\bibinfo {volume} {60}},\ \bibinfo {pages} {439} (\bibinfo {year} {2010})},\
  \Eprint {http://arxiv.org/abs/1004.3311} {arXiv:1004.3311 [astro-ph.HE]}
  \BibitemShut {NoStop}%
%%CITATION = ARXIV:1004.3311;%%
\bibitem [{\citenamefont {Taylor}(1975)}]{Moondata}%
  \BibitemOpen
  \bibfield  {author} {\bibinfo {author} {\bibfnamefont {S.~R.}\ \bibnamefont
  {Taylor}},\ }\href@noop {} {\emph {\bibinfo {title} {Lunar science: a
  post-Apollo view; scientific results and insights from the lunar samples}}}\
  (\bibinfo  {publisher} {Pergamon Press},\ \bibinfo {year} {1975})\BibitemShut
  {NoStop}%
\bibitem [{\citenamefont {Brun}\ \emph {et~al.}(1994)\citenamefont {Brun},
  \citenamefont {Bruyant}, \citenamefont {Carminati}, \citenamefont {Giani},
  \citenamefont {Maire}, \citenamefont {McPherson}, \citenamefont {Patrick},\
  and\ \citenamefont {Urban}}]{Brun:1994aa}%
  \BibitemOpen
  \bibfield  {author} {\bibinfo {author} {\bibfnamefont {R.}~\bibnamefont
  {Brun}}, \bibinfo {author} {\bibfnamefont {F.}~\bibnamefont {Bruyant}},
  \bibinfo {author} {\bibfnamefont {F.}~\bibnamefont {Carminati}}, \bibinfo
  {author} {\bibfnamefont {S.}~\bibnamefont {Giani}}, \bibinfo {author}
  {\bibfnamefont {M.}~\bibnamefont {Maire}}, \bibinfo {author} {\bibfnamefont
  {A.}~\bibnamefont {McPherson}}, \bibinfo {author} {\bibfnamefont
  {G.}~\bibnamefont {Patrick}}, \ and\ \bibinfo {author} {\bibfnamefont
  {L.}~\bibnamefont {Urban}},\ }\href {\doibase 10.17181/CERN.MUHF.DMJ1} {\
  (\bibinfo {year} {1994}),\ 10.17181/CERN.MUHF.DMJ1}\BibitemShut {NoStop}%
\bibitem [{\citenamefont {Agostinelli}\ \emph {et~al.}(2003)\citenamefont
  {Agostinelli} \emph {et~al.}}]{GEANT4:2002zbu}%
  \BibitemOpen
  \bibfield  {author} {\bibinfo {author} {\bibfnamefont {S.}~\bibnamefont
  {Agostinelli}} \emph {et~al.} (\bibinfo {collaboration} {GEANT4}),\ }\href
  {\doibase 10.1016/S0168-9002(03)01368-8} {\bibfield  {journal} {\bibinfo
  {journal} {Nucl. Instrum. Meth. A}\ }\textbf {\bibinfo {volume} {506}},\
  \bibinfo {pages} {250} (\bibinfo {year} {2003})}\BibitemShut {NoStop}%
\bibitem [{\citenamefont {Zyla}\ \emph {et~al.}(2020)\citenamefont {Zyla} \emph
  {et~al.}}]{Zyla:2020zbs}%
  \BibitemOpen
  \bibfield  {author} {\bibinfo {author} {\bibfnamefont {P.~A.}\ \bibnamefont
  {Zyla}} \emph {et~al.} (\bibinfo {collaboration} {Particle Data Group}),\
  }\href {\doibase 10.1093/ptep/ptaa104} {\bibfield  {journal} {\bibinfo
  {journal} {PTEP}\ }\textbf {\bibinfo {volume} {2020}},\ \bibinfo {pages}
  {083C01} (\bibinfo {year} {2020})}\BibitemShut {NoStop}%
%%CITATION = INSPIRE-1812251;%%
\bibitem [{\citenamefont {Ferrari}\ \emph {et~al.}(2005)\citenamefont
  {Ferrari}, \citenamefont {Sala}, \citenamefont {Fasso},\ and\ \citenamefont
  {Ranft}}]{Ferrari:2005zk}%
  \BibitemOpen
  \bibfield  {author} {\bibinfo {author} {\bibfnamefont {A.}~\bibnamefont
  {Ferrari}}, \bibinfo {author} {\bibfnamefont {P.~R.}\ \bibnamefont {Sala}},
  \bibinfo {author} {\bibfnamefont {A.}~\bibnamefont {Fasso}}, \ and\ \bibinfo
  {author} {\bibfnamefont {J.}~\bibnamefont {Ranft}},\ }\href {\doibase
  10.2172/877507} {\  (\bibinfo {year} {2005}),\ 10.2172/877507}\BibitemShut
  {NoStop}%
\bibitem [{\citenamefont {Gleeson}\ and\ \citenamefont
  {Axford}(1968)}]{Gleeson:1968zza}%
  \BibitemOpen
  \bibfield  {author} {\bibinfo {author} {\bibfnamefont {L.~J.}\ \bibnamefont
  {Gleeson}}\ and\ \bibinfo {author} {\bibfnamefont {W.~I.}\ \bibnamefont
  {Axford}},\ }\href {\doibase 10.1086/149822} {\bibfield  {journal} {\bibinfo
  {journal} {Astrophys. J.}\ }\textbf {\bibinfo {volume} {154}},\ \bibinfo
  {pages} {1011} (\bibinfo {year} {1968})}\BibitemShut {NoStop}%
\bibitem [{\citenamefont {Potgieter}(2013)}]{Potgieter:2013pdj}%
  \BibitemOpen
  \bibfield  {author} {\bibinfo {author} {\bibfnamefont {M.}~\bibnamefont
  {Potgieter}},\ }\href {\doibase 10.12942/lrsp-2013-3} {\bibfield  {journal}
  {\bibinfo  {journal} {Living Rev. Solar Phys.}\ }\textbf {\bibinfo {volume}
  {10}},\ \bibinfo {pages} {3} (\bibinfo {year} {2013})},\ \Eprint
  {http://arxiv.org/abs/1306.4421} {arXiv:1306.4421 [physics.space-ph]}
  \BibitemShut {NoStop}%
\bibitem [{\citenamefont {Gelmini}\ \emph {et~al.}(2019)\citenamefont
  {Gelmini}, \citenamefont {Takhistov},\ and\ \citenamefont
  {Witte}}]{Gelmini:2018gqa}%
  \BibitemOpen
  \bibfield  {author} {\bibinfo {author} {\bibfnamefont {G.~B.}\ \bibnamefont
  {Gelmini}}, \bibinfo {author} {\bibfnamefont {V.}~\bibnamefont {Takhistov}},
  \ and\ \bibinfo {author} {\bibfnamefont {S.~J.}\ \bibnamefont {Witte}},\
  }\href {\doibase 10.1103/PhysRevD.99.093009} {\bibfield  {journal} {\bibinfo
  {journal} {Phys. Rev. D}\ }\textbf {\bibinfo {volume} {99}},\ \bibinfo
  {pages} {093009} (\bibinfo {year} {2019})},\ \Eprint
  {http://arxiv.org/abs/1812.05550} {arXiv:1812.05550 [hep-ph]} \BibitemShut
  {NoStop}%
\bibitem [{\citenamefont {Reames}(2021)}]{Reames:2020vxe}%
  \BibitemOpen
  \bibfield  {author} {\bibinfo {author} {\bibfnamefont {D.~V.}\ \bibnamefont
  {Reames}},\ }\href {\doibase 10.1007/978-3-030-66402-2} {\bibfield  {journal}
  {\bibinfo  {journal} {Lect. Notes Phys.}\ }\textbf {\bibinfo {volume}
  {978}},\ \bibinfo {pages} {pp.} (\bibinfo {year} {2021})},\ \Eprint
  {http://arxiv.org/abs/2010.08517} {arXiv:2010.08517 [astro-ph.SR]}
  \BibitemShut {NoStop}%
\bibitem [{\citenamefont {Bruno}\ \emph {et~al.}(2018)\citenamefont {Bruno}
  \emph {et~al.}}]{PAMELA:2018yie}%
  \BibitemOpen
  \bibfield  {author} {\bibinfo {author} {\bibfnamefont {A.}~\bibnamefont
  {Bruno}} \emph {et~al.} (\bibinfo {collaboration} {PAMELA}),\ }\href
  {\doibase 10.3847/1538-4357/aacc26} {\bibfield  {journal} {\bibinfo
  {journal} {Astrophys. J.}\ }\textbf {\bibinfo {volume} {862}},\ \bibinfo
  {pages} {97} (\bibinfo {year} {2018})},\ \Eprint
  {http://arxiv.org/abs/1807.10183} {arXiv:1807.10183 [astro-ph.SR]}
  \BibitemShut {NoStop}%
\bibitem [{\citenamefont {Freedman}\ \emph {et~al.}(1977)\citenamefont
  {Freedman}, \citenamefont {Schramm},\ and\ \citenamefont
  {Tubbs}}]{Freedman:1977xn}%
  \BibitemOpen
  \bibfield  {author} {\bibinfo {author} {\bibfnamefont {D.~Z.}\ \bibnamefont
  {Freedman}}, \bibinfo {author} {\bibfnamefont {D.~N.}\ \bibnamefont
  {Schramm}}, \ and\ \bibinfo {author} {\bibfnamefont {D.~L.}\ \bibnamefont
  {Tubbs}},\ }\href {\doibase 10.1146/annurev.ns.27.120177.001123} {\bibfield
  {journal} {\bibinfo  {journal} {Ann. Rev. Nucl. Part. Sci.}\ }\textbf
  {\bibinfo {volume} {27}},\ \bibinfo {pages} {167} (\bibinfo {year}
  {1977})}\BibitemShut {NoStop}%
%%CITATION = ARNUA,27,167;%%
\bibitem [{\citenamefont {Szydagis}\ \emph {et~al.}(2021)\citenamefont
  {Szydagis} \emph {et~al.}}]{Szydagis:2021hfh}%
  \BibitemOpen
  \bibfield  {author} {\bibinfo {author} {\bibfnamefont {M.}~\bibnamefont
  {Szydagis}} \emph {et~al.},\ }\href {\doibase 10.3390/instruments5010013}
  {\bibfield  {journal} {\bibinfo  {journal} {Instruments}\ }\textbf {\bibinfo
  {volume} {5}},\ \bibinfo {pages} {13} (\bibinfo {year} {2021})},\ \Eprint
  {http://arxiv.org/abs/2102.10209} {arXiv:2102.10209 [hep-ex]} \BibitemShut
  {NoStop}%
%%CITATION = ARXIV:2102.10209;%%
\bibitem [{\citenamefont {Sarantakos}\ \emph {et~al.}(1983)\citenamefont
  {Sarantakos}, \citenamefont {Sirlin},\ and\ \citenamefont
  {Marciano}}]{Sarantakos:1982bp}%
  \BibitemOpen
  \bibfield  {author} {\bibinfo {author} {\bibfnamefont {S.}~\bibnamefont
  {Sarantakos}}, \bibinfo {author} {\bibfnamefont {A.}~\bibnamefont {Sirlin}},
  \ and\ \bibinfo {author} {\bibfnamefont {W.~J.}\ \bibnamefont {Marciano}},\
  }\href {\doibase 10.1016/0550-3213(83)90079-2} {\bibfield  {journal}
  {\bibinfo  {journal} {Nucl. Phys.}\ }\textbf {\bibinfo {volume} {B217}},\
  \bibinfo {pages} {84} (\bibinfo {year} {1983})}\BibitemShut {NoStop}%
%%CITATION = NUPHA,B217,84;%%
\bibitem [{\citenamefont {Vogel}\ and\ \citenamefont
  {Engel}(1989)}]{Vogel:1989iv}%
  \BibitemOpen
  \bibfield  {author} {\bibinfo {author} {\bibfnamefont {P.}~\bibnamefont
  {Vogel}}\ and\ \bibinfo {author} {\bibfnamefont {J.}~\bibnamefont {Engel}},\
  }\href {\doibase 10.1103/PhysRevD.39.3378} {\bibfield  {journal} {\bibinfo
  {journal} {Phys. Rev.}\ }\textbf {\bibinfo {volume} {D39}},\ \bibinfo {pages}
  {3378} (\bibinfo {year} {1989})}\BibitemShut {NoStop}%
%%CITATION = PHRVA,D39,3378;%%
\bibitem [{\citenamefont {Marciano}\ and\ \citenamefont
  {Parsa}(2003)}]{Marciano:2003eq}%
  \BibitemOpen
  \bibfield  {author} {\bibinfo {author} {\bibfnamefont {W.~J.}\ \bibnamefont
  {Marciano}}\ and\ \bibinfo {author} {\bibfnamefont {Z.}~\bibnamefont
  {Parsa}},\ }\href {\doibase 10.1088/0954-3899/29/11/013} {\bibfield
  {journal} {\bibinfo  {journal} {J. Phys.}\ }\textbf {\bibinfo {volume}
  {G29}},\ \bibinfo {pages} {2629} (\bibinfo {year} {2003})},\ \Eprint
  {http://arxiv.org/abs/hep-ph/0403168} {arXiv:hep-ph/0403168 [hep-ph]}
  \BibitemShut {NoStop}%
%%CITATION = HEP-PH/0403168;%%
\bibitem [{\citenamefont {Chen}\ \emph {et~al.}(2017)\citenamefont {Chen},
  \citenamefont {Chi}, \citenamefont {Liu},\ and\ \citenamefont
  {Wu}}]{Chen:2016eab}%
  \BibitemOpen
  \bibfield  {author} {\bibinfo {author} {\bibfnamefont {J.-W.}\ \bibnamefont
  {Chen}}, \bibinfo {author} {\bibfnamefont {H.-C.}\ \bibnamefont {Chi}},
  \bibinfo {author} {\bibfnamefont {C.~P.}\ \bibnamefont {Liu}}, \ and\
  \bibinfo {author} {\bibfnamefont {C.-P.}\ \bibnamefont {Wu}},\ }\href
  {\doibase 10.1016/j.physletb.2017.10.029} {\bibfield  {journal} {\bibinfo
  {journal} {Phys. Lett.}\ }\textbf {\bibinfo {volume} {B774}},\ \bibinfo
  {pages} {656} (\bibinfo {year} {2017})},\ \Eprint
  {http://arxiv.org/abs/1610.04177} {arXiv:1610.04177 [hep-ex]} \BibitemShut
  {NoStop}%
%%CITATION = ARXIV:1610.04177;%%
\bibitem [{\citenamefont {Baxter}\ \emph {et~al.}(2021)\citenamefont {Baxter}
  \emph {et~al.}}]{Baxter:2021pqo}%
  \BibitemOpen
  \bibfield  {author} {\bibinfo {author} {\bibfnamefont {D.}~\bibnamefont
  {Baxter}} \emph {et~al.},\ }\href {\doibase 10.1140/epjc/s10052-021-09655-y}
  {\bibfield  {journal} {\bibinfo  {journal} {Eur. Phys. J. C}\ }\textbf
  {\bibinfo {volume} {81}},\ \bibinfo {pages} {907} (\bibinfo {year} {2021})},\
  \Eprint {http://arxiv.org/abs/2105.00599} {arXiv:2105.00599 [hep-ex]}
  \BibitemShut {NoStop}%
\bibitem [{\citenamefont {Grevesse}\ and\ \citenamefont
  {Sauval}(1998)}]{Grevesse:1998bj}%
  \BibitemOpen
  \bibfield  {author} {\bibinfo {author} {\bibfnamefont {N.}~\bibnamefont
  {Grevesse}}\ and\ \bibinfo {author} {\bibfnamefont {A.~J.}\ \bibnamefont
  {Sauval}},\ }\href {\doibase 10.1023/A:1005161325181} {\bibfield  {journal}
  {\bibinfo  {journal} {Space Sci. Rev.}\ }\textbf {\bibinfo {volume} {85}},\
  \bibinfo {pages} {161} (\bibinfo {year} {1998})}\BibitemShut {NoStop}%
%%CITATION = SPSRA,85,161;%%
\bibitem [{\citenamefont {Asplund}\ \emph {et~al.}(2009)\citenamefont
  {Asplund}, \citenamefont {Grevesse}, \citenamefont {Sauval},\ and\
  \citenamefont {Scott}}]{Asplund:2009fu}%
  \BibitemOpen
  \bibfield  {author} {\bibinfo {author} {\bibfnamefont {M.}~\bibnamefont
  {Asplund}}, \bibinfo {author} {\bibfnamefont {N.}~\bibnamefont {Grevesse}},
  \bibinfo {author} {\bibfnamefont {A.~J.}\ \bibnamefont {Sauval}}, \ and\
  \bibinfo {author} {\bibfnamefont {P.}~\bibnamefont {Scott}},\ }\href
  {\doibase 10.1146/annurev.astro.46.060407.145222} {\bibfield  {journal}
  {\bibinfo  {journal} {Ann. Rev. Astron. Astrophys.}\ }\textbf {\bibinfo
  {volume} {47}},\ \bibinfo {pages} {481} (\bibinfo {year} {2009})},\ \Eprint
  {http://arxiv.org/abs/0909.0948} {arXiv:0909.0948 [astro-ph.SR]} \BibitemShut
  {NoStop}%
%%CITATION = ARXIV:0909.0948;%%
\bibitem [{\citenamefont {Serenelli}\ \emph {et~al.}(2011)\citenamefont
  {Serenelli}, \citenamefont {Haxton},\ and\ \citenamefont
  {Pena-Garay}}]{Serenelli:2011py}%
  \BibitemOpen
  \bibfield  {author} {\bibinfo {author} {\bibfnamefont {A.~M.}\ \bibnamefont
  {Serenelli}}, \bibinfo {author} {\bibfnamefont {W.~C.}\ \bibnamefont
  {Haxton}}, \ and\ \bibinfo {author} {\bibfnamefont {C.}~\bibnamefont
  {Pena-Garay}},\ }\href {\doibase 10.1088/0004-637X/743/1/24} {\bibfield
  {journal} {\bibinfo  {journal} {Astrophys. J.}\ }\textbf {\bibinfo {volume}
  {743}},\ \bibinfo {pages} {24} (\bibinfo {year} {2011})},\ \Eprint
  {http://arxiv.org/abs/1104.1639} {arXiv:1104.1639 [astro-ph.SR]} \BibitemShut
  {NoStop}%
%%CITATION = ARXIV:1104.1639;%%
\bibitem [{\citenamefont {Bergstrom}\ \emph {et~al.}(2016)\citenamefont
  {Bergstrom}, \citenamefont {Gonzalez-Garcia}, \citenamefont {Maltoni},
  \citenamefont {Pena-Garay}, \citenamefont {Serenelli},\ and\ \citenamefont
  {Song}}]{Bergstrom:2016cbh}%
  \BibitemOpen
  \bibfield  {author} {\bibinfo {author} {\bibfnamefont {J.}~\bibnamefont
  {Bergstrom}}, \bibinfo {author} {\bibfnamefont {M.~C.}\ \bibnamefont
  {Gonzalez-Garcia}}, \bibinfo {author} {\bibfnamefont {M.}~\bibnamefont
  {Maltoni}}, \bibinfo {author} {\bibfnamefont {C.}~\bibnamefont {Pena-Garay}},
  \bibinfo {author} {\bibfnamefont {A.~M.}\ \bibnamefont {Serenelli}}, \ and\
  \bibinfo {author} {\bibfnamefont {N.}~\bibnamefont {Song}},\ }\href {\doibase
  10.1007/JHEP03(2016)132} {\bibfield  {journal} {\bibinfo  {journal} {JHEP}\
  }\textbf {\bibinfo {volume} {03}},\ \bibinfo {pages} {132} (\bibinfo {year}
  {2016})},\ \Eprint {http://arxiv.org/abs/1601.00972} {arXiv:1601.00972
  [hep-ph]} \BibitemShut {NoStop}%
%%CITATION = ARXIV:1601.00972;%%
\bibitem [{\citenamefont {Agostini}\ \emph {et~al.}(2020)\citenamefont
  {Agostini} \emph {et~al.}}]{Agostini:2020mfq}%
  \BibitemOpen
  \bibfield  {author} {\bibinfo {author} {\bibfnamefont {M.}~\bibnamefont
  {Agostini}} \emph {et~al.} (\bibinfo {collaboration} {BOREXINO}),\ }\href
  {\doibase 10.1038/s41586-020-2934-0} {\bibfield  {journal} {\bibinfo
  {journal} {Nature}\ }\textbf {\bibinfo {volume} {587}},\ \bibinfo {pages}
  {577} (\bibinfo {year} {2020})},\ \Eprint {http://arxiv.org/abs/2006.15115}
  {arXiv:2006.15115 [hep-ex]} \BibitemShut {NoStop}%
%%CITATION = ARXIV:2006.15115;%%
\bibitem [{\citenamefont {Battistoni}\ \emph {et~al.}(2005)\citenamefont
  {Battistoni}, \citenamefont {Ferrari}, \citenamefont {Montaruli},\ and\
  \citenamefont {Sala}}]{Battistoni:2005pd}%
  \BibitemOpen
  \bibfield  {author} {\bibinfo {author} {\bibfnamefont {G.}~\bibnamefont
  {Battistoni}}, \bibinfo {author} {\bibfnamefont {A.}~\bibnamefont {Ferrari}},
  \bibinfo {author} {\bibfnamefont {T.}~\bibnamefont {Montaruli}}, \ and\
  \bibinfo {author} {\bibfnamefont {P.~R.}\ \bibnamefont {Sala}},\ }\href
  {\doibase 10.1016/j.astropartphys.2005.03.006} {\bibfield  {journal}
  {\bibinfo  {journal} {Astropart. Phys.}\ }\textbf {\bibinfo {volume} {23}},\
  \bibinfo {pages} {526} (\bibinfo {year} {2005})}\BibitemShut {NoStop}%
%%CITATION = APHYE,23,526;%%
\bibitem [{\citenamefont {Barr}\ \emph {et~al.}(2006)\citenamefont {Barr},
  \citenamefont {Gaisser}, \citenamefont {Robbins},\ and\ \citenamefont
  {Stanev}}]{Barr:2006it}%
  \BibitemOpen
  \bibfield  {author} {\bibinfo {author} {\bibfnamefont {G.~D.}\ \bibnamefont
  {Barr}}, \bibinfo {author} {\bibfnamefont {T.~K.}\ \bibnamefont {Gaisser}},
  \bibinfo {author} {\bibfnamefont {S.}~\bibnamefont {Robbins}}, \ and\
  \bibinfo {author} {\bibfnamefont {T.}~\bibnamefont {Stanev}},\ }\href
  {\doibase 10.1103/PhysRevD.74.094009} {\bibfield  {journal} {\bibinfo
  {journal} {Phys. Rev.}\ }\textbf {\bibinfo {volume} {D74}},\ \bibinfo {pages}
  {094009} (\bibinfo {year} {2006})},\ \Eprint
  {http://arxiv.org/abs/astro-ph/0611266} {arXiv:astro-ph/0611266 [astro-ph]}
  \BibitemShut {NoStop}%
%%CITATION = ASTRO-PH/0611266;%%
\bibitem [{\citenamefont {Honda}\ \emph {et~al.}(2011)\citenamefont {Honda},
  \citenamefont {Kajita}, \citenamefont {Kasahara},\ and\ \citenamefont
  {Midorikawa}}]{Honda:2011nf}%
  \BibitemOpen
  \bibfield  {author} {\bibinfo {author} {\bibfnamefont {M.}~\bibnamefont
  {Honda}}, \bibinfo {author} {\bibfnamefont {T.}~\bibnamefont {Kajita}},
  \bibinfo {author} {\bibfnamefont {K.}~\bibnamefont {Kasahara}}, \ and\
  \bibinfo {author} {\bibfnamefont {S.}~\bibnamefont {Midorikawa}},\ }\href
  {\doibase 10.1103/PhysRevD.83.123001} {\bibfield  {journal} {\bibinfo
  {journal} {Phys. Rev.}\ }\textbf {\bibinfo {volume} {D83}},\ \bibinfo {pages}
  {123001} (\bibinfo {year} {2011})},\ \Eprint {http://arxiv.org/abs/1102.2688}
  {arXiv:1102.2688 [astro-ph.HE]} \BibitemShut {NoStop}%
%%CITATION = ARXIV:1102.2688;%%
\bibitem [{\citenamefont {Honda}\ \emph {et~al.}(2015)\citenamefont {Honda},
  \citenamefont {Sajjad~Athar}, \citenamefont {Kajita}, \citenamefont
  {Kasahara},\ and\ \citenamefont {Midorikawa}}]{Honda:2015fha}%
  \BibitemOpen
  \bibfield  {author} {\bibinfo {author} {\bibfnamefont {M.}~\bibnamefont
  {Honda}}, \bibinfo {author} {\bibfnamefont {M.}~\bibnamefont {Sajjad~Athar}},
  \bibinfo {author} {\bibfnamefont {T.}~\bibnamefont {Kajita}}, \bibinfo
  {author} {\bibfnamefont {K.}~\bibnamefont {Kasahara}}, \ and\ \bibinfo
  {author} {\bibfnamefont {S.}~\bibnamefont {Midorikawa}},\ }\href {\doibase
  10.1103/PhysRevD.92.023004} {\bibfield  {journal} {\bibinfo  {journal} {Phys.
  Rev.}\ }\textbf {\bibinfo {volume} {D92}},\ \bibinfo {pages} {023004}
  (\bibinfo {year} {2015})},\ \Eprint {http://arxiv.org/abs/1502.03916}
  {arXiv:1502.03916 [astro-ph.HE]} \BibitemShut {NoStop}%
%%CITATION = ARXIV:1502.03916;%%
\bibitem [{\citenamefont {Cowan}\ \emph {et~al.}(2011)\citenamefont {Cowan},
  \citenamefont {Cranmer}, \citenamefont {Gross},\ and\ \citenamefont
  {Vitells}}]{Cowan:2010js}%
  \BibitemOpen
  \bibfield  {author} {\bibinfo {author} {\bibfnamefont {G.}~\bibnamefont
  {Cowan}}, \bibinfo {author} {\bibfnamefont {K.}~\bibnamefont {Cranmer}},
  \bibinfo {author} {\bibfnamefont {E.}~\bibnamefont {Gross}}, \ and\ \bibinfo
  {author} {\bibfnamefont {O.}~\bibnamefont {Vitells}},\ }\href {\doibase
  10.1140/epjc/s10052-011-1554-0, 10.1140/epjc/s10052-013-2501-z} {\bibfield
  {journal} {\bibinfo  {journal} {Eur. Phys. J.}\ }\textbf {\bibinfo {volume}
  {C71}},\ \bibinfo {pages} {1554} (\bibinfo {year} {2011})},\ \bibinfo {note}
  {[Erratum: Eur. Phys. J.C73,2501(2013)]},\ \Eprint
  {http://arxiv.org/abs/1007.1727} {arXiv:1007.1727 [physics.data-an]}
  \BibitemShut {NoStop}%
%%CITATION = ARXIV:1007.1727;%%
\bibitem [{\citenamefont {Tang}\ and\ \citenamefont
  {Zhang}(2023)}]{Tang:2023xub}%
  \BibitemOpen
  \bibfield  {author} {\bibinfo {author} {\bibfnamefont {J.}~\bibnamefont
  {Tang}}\ and\ \bibinfo {author} {\bibfnamefont {B.-L.}\ \bibnamefont
  {Zhang}},\ }\href@noop {} {\  (\bibinfo {year} {2023})},\ \Eprint
  {http://arxiv.org/abs/2304.13665} {arXiv:2304.13665 [hep-ph]} \BibitemShut
  {NoStop}%
\bibitem [{\citenamefont {Boulay}\ and\ \citenamefont
  {Hime}(2006)}]{Boulay:2006mb}%
  \BibitemOpen
  \bibfield  {author} {\bibinfo {author} {\bibfnamefont {M.~G.}\ \bibnamefont
  {Boulay}}\ and\ \bibinfo {author} {\bibfnamefont {A.}~\bibnamefont {Hime}},\
  }\href {\doibase 10.1016/j.astropartphys.2005.12.009} {\bibfield  {journal}
  {\bibinfo  {journal} {Astropart. Phys.}\ }\textbf {\bibinfo {volume} {25}},\
  \bibinfo {pages} {179} (\bibinfo {year} {2006})}\BibitemShut {NoStop}%
%%CITATION = APHYE,25,179;%%
\bibitem [{\citenamefont {Amaudruz}\ \emph {et~al.}(2016)\citenamefont
  {Amaudruz} \emph {et~al.}}]{DEAP:2009hyz}%
  \BibitemOpen
  \bibfield  {author} {\bibinfo {author} {\bibfnamefont {P.~A.}\ \bibnamefont
  {Amaudruz}} \emph {et~al.} (\bibinfo {collaboration} {DEAP}),\ }\href
  {\doibase 10.1016/j.astropartphys.2016.09.002} {\bibfield  {journal}
  {\bibinfo  {journal} {Astropart. Phys.}\ }\textbf {\bibinfo {volume} {85}},\
  \bibinfo {pages} {1} (\bibinfo {year} {2016})},\ \Eprint
  {http://arxiv.org/abs/0904.2930} {arXiv:0904.2930 [astro-ph.IM]} \BibitemShut
  {NoStop}%
\bibitem [{\citenamefont {Adhikari}\ \emph {et~al.}(2020)\citenamefont
  {Adhikari} \emph {et~al.}}]{Adhikari:2020zyy}%
  \BibitemOpen
  \bibfield  {author} {\bibinfo {author} {\bibfnamefont {P.}~\bibnamefont
  {Adhikari}} \emph {et~al.} (\bibinfo {collaboration} {DEAP}),\ }\href
  {\doibase 10.1140/epjc/s10052-020-7789-x} {\bibfield  {journal} {\bibinfo
  {journal} {Eur. Phys. J. C}\ }\textbf {\bibinfo {volume} {80}},\ \bibinfo
  {pages} {303} (\bibinfo {year} {2020})},\ \Eprint
  {http://arxiv.org/abs/2001.09855} {arXiv:2001.09855 [physics.ins-det]}
  \BibitemShut {NoStop}%
\bibitem [{\citenamefont {Adhikari}\ \emph {et~al.}(2021)\citenamefont
  {Adhikari} \emph {et~al.}}]{DEAP:2021axq}%
  \BibitemOpen
  \bibfield  {author} {\bibinfo {author} {\bibfnamefont {P.}~\bibnamefont
  {Adhikari}} \emph {et~al.} (\bibinfo {collaboration} {DEAP}),\ }\href
  {\doibase 10.1140/epjc/s10052-021-09514-w} {\bibfield  {journal} {\bibinfo
  {journal} {Eur. Phys. J. C}\ }\textbf {\bibinfo {volume} {81}},\ \bibinfo
  {pages} {823} (\bibinfo {year} {2021})},\ \Eprint
  {http://arxiv.org/abs/2103.12202} {arXiv:2103.12202 [physics.ins-det]}
  \BibitemShut {NoStop}%
\bibitem [{\citenamefont {Wilks}(1938)}]{Wilks:1938dza}%
  \BibitemOpen
  \bibfield  {author} {\bibinfo {author} {\bibfnamefont {S.~S.}\ \bibnamefont
  {Wilks}},\ }\href {\doibase 10.1214/aoms/1177732360} {\bibfield  {journal}
  {\bibinfo  {journal} {Annals Math. Statist.}\ }\textbf {\bibinfo {volume}
  {9}},\ \bibinfo {pages} {60} (\bibinfo {year} {1938})}\BibitemShut {NoStop}%
%%CITATION = AASTA,9,60;%%
\bibitem [{\citenamefont {Wald}(1943)}]{Wald:1942}%
  \BibitemOpen
  \bibfield  {author} {\bibinfo {author} {\bibfnamefont {A.}~\bibnamefont
  {Wald}},\ }\href {\doibase 10.2307/1990256} {\bibfield  {journal} {\bibinfo
  {journal} {Trans. Am. Math. Soc.}\ }\textbf {\bibinfo {volume} {54}},\
  \bibinfo {pages} {426} (\bibinfo {year} {1943})}\BibitemShut {NoStop}%
\bibitem [{\citenamefont {Engel}\ \emph {et~al.}(1992)\citenamefont {Engel},
  \citenamefont {Pittel},\ and\ \citenamefont {Vogel}}]{Engel:1992bf}%
  \BibitemOpen
  \bibfield  {author} {\bibinfo {author} {\bibfnamefont {J.}~\bibnamefont
  {Engel}}, \bibinfo {author} {\bibfnamefont {S.}~\bibnamefont {Pittel}}, \
  and\ \bibinfo {author} {\bibfnamefont {P.}~\bibnamefont {Vogel}},\ }\href
  {\doibase 10.1142/S0218301392000023} {\bibfield  {journal} {\bibinfo
  {journal} {Int. J. Mod. Phys.}\ }\textbf {\bibinfo {volume} {E1}},\ \bibinfo
  {pages} {1} (\bibinfo {year} {1992})}\BibitemShut {NoStop}%
%%CITATION = IMPAE,E1,1;%%
\bibitem [{\citenamefont {Formaggio}\ and\ \citenamefont
  {Zeller}(2012)}]{Formaggio:2013kya}%
  \BibitemOpen
  \bibfield  {author} {\bibinfo {author} {\bibfnamefont {J.~A.}\ \bibnamefont
  {Formaggio}}\ and\ \bibinfo {author} {\bibfnamefont {G.~P.}\ \bibnamefont
  {Zeller}},\ }\href {\doibase 10.1103/RevModPhys.84.1307} {\bibfield
  {journal} {\bibinfo  {journal} {Rev. Mod. Phys.}\ }\textbf {\bibinfo {volume}
  {84}},\ \bibinfo {pages} {1307} (\bibinfo {year} {2012})},\ \Eprint
  {http://arxiv.org/abs/1305.7513} {arXiv:1305.7513 [hep-ex]} \BibitemShut
  {NoStop}%
%%CITATION = ARXIV:1305.7513;%%
\bibitem [{\citenamefont {Voloshin}(2010)}]{Voloshin:2010vm}%
  \BibitemOpen
  \bibfield  {author} {\bibinfo {author} {\bibfnamefont {M.~B.}\ \bibnamefont
  {Voloshin}},\ }\href {\doibase 10.1103/PhysRevLett.105.201801,
  10.1103/PhysRevLett.106.059901} {\bibfield  {journal} {\bibinfo  {journal}
  {Phys. Rev. Lett.}\ }\textbf {\bibinfo {volume} {105}},\ \bibinfo {pages}
  {201801} (\bibinfo {year} {2010})},\ \bibinfo {note} {[Erratum: Phys. Rev.
  Lett.106,059901(2011)]},\ \Eprint {http://arxiv.org/abs/1008.2171}
  {arXiv:1008.2171 [hep-ph]} \BibitemShut {NoStop}%
%%CITATION = ARXIV:1008.2171;%%
\bibitem [{\citenamefont {Kouzakov}\ \emph
  {et~al.}(2011{\natexlab{a}})\citenamefont {Kouzakov}, \citenamefont
  {Studenikin},\ and\ \citenamefont {Voloshin}}]{Kouzakov:2011ka}%
  \BibitemOpen
  \bibfield  {author} {\bibinfo {author} {\bibfnamefont {K.~A.}\ \bibnamefont
  {Kouzakov}}, \bibinfo {author} {\bibfnamefont {A.~I.}\ \bibnamefont
  {Studenikin}}, \ and\ \bibinfo {author} {\bibfnamefont {M.~B.}\ \bibnamefont
  {Voloshin}},\ }\href {\doibase 10.1134/S0021364011110075} {\bibfield
  {journal} {\bibinfo  {journal} {JETP Lett.}\ }\textbf {\bibinfo {volume}
  {93}},\ \bibinfo {pages} {623} (\bibinfo {year} {2011}{\natexlab{a}})},\
  \Eprint {http://arxiv.org/abs/1105.5543} {arXiv:1105.5543 [hep-ph]}
  \BibitemShut {NoStop}%
%%CITATION = ARXIV:1105.5543;%%
\bibitem [{\citenamefont {Kouzakov}\ \emph
  {et~al.}(2011{\natexlab{b}})\citenamefont {Kouzakov}, \citenamefont
  {Studenikin},\ and\ \citenamefont {Voloshin}}]{Kouzakov:2011vx}%
  \BibitemOpen
  \bibfield  {author} {\bibinfo {author} {\bibfnamefont {K.~A.}\ \bibnamefont
  {Kouzakov}}, \bibinfo {author} {\bibfnamefont {A.~I.}\ \bibnamefont
  {Studenikin}}, \ and\ \bibinfo {author} {\bibfnamefont {M.~B.}\ \bibnamefont
  {Voloshin}},\ }\href {\doibase 10.1103/PhysRevD.83.113001} {\bibfield
  {journal} {\bibinfo  {journal} {Phys. Rev.}\ }\textbf {\bibinfo {volume}
  {D83}},\ \bibinfo {pages} {113001} (\bibinfo {year} {2011}{\natexlab{b}})},\
  \Eprint {http://arxiv.org/abs/1101.4878} {arXiv:1101.4878 [hep-ph]}
  \BibitemShut {NoStop}%
%%CITATION = ARXIV:1101.4878;%%
\bibitem [{\citenamefont {Kouzakov}\ and\ \citenamefont
  {Studenikin}(2014)}]{Kouzakov:2014lka}%
  \BibitemOpen
  \bibfield  {author} {\bibinfo {author} {\bibfnamefont {K.~A.}\ \bibnamefont
  {Kouzakov}}\ and\ \bibinfo {author} {\bibfnamefont {A.~I.}\ \bibnamefont
  {Studenikin}},\ }\href {\doibase 10.1155/2014/569409} {\bibfield  {journal}
  {\bibinfo  {journal} {Adv. High Energy Phys.}\ }\textbf {\bibinfo {volume}
  {2014}},\ \bibinfo {pages} {569409} (\bibinfo {year} {2014})},\ \Eprint
  {http://arxiv.org/abs/1406.4999} {arXiv:1406.4999 [hep-ph]} \BibitemShut
  {NoStop}%
%%CITATION = ARXIV:1406.4999;%%
\end{thebibliography}%

%%%%%%%%%%%%%%%%%%%%%%%%%%%%%%%%%%%%%%

\clearpage
\onecolumngrid
\appendix

\section{Supplemental Material\label{sec:sup}}

In this supplemental material we present further details underlying the calculations and analysis in the main text.

\subsection{More on Cosmic Ray Neutrino Fluxes on the Moon}

\begin{figure}[ttt]
  \begin{center}
    \includegraphics[width = 0.47\textwidth]{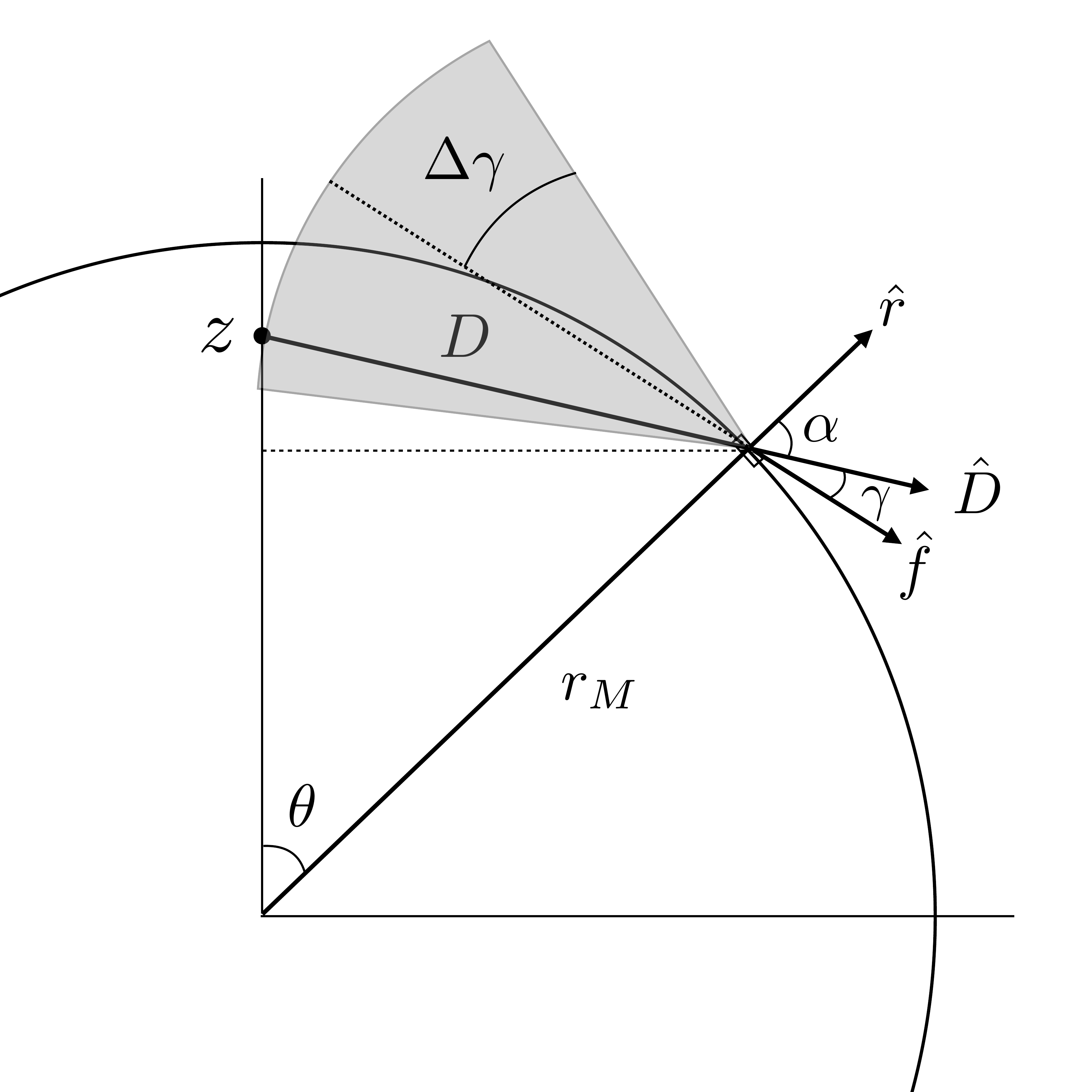}
    \includegraphics[width = 0.45\textwidth]{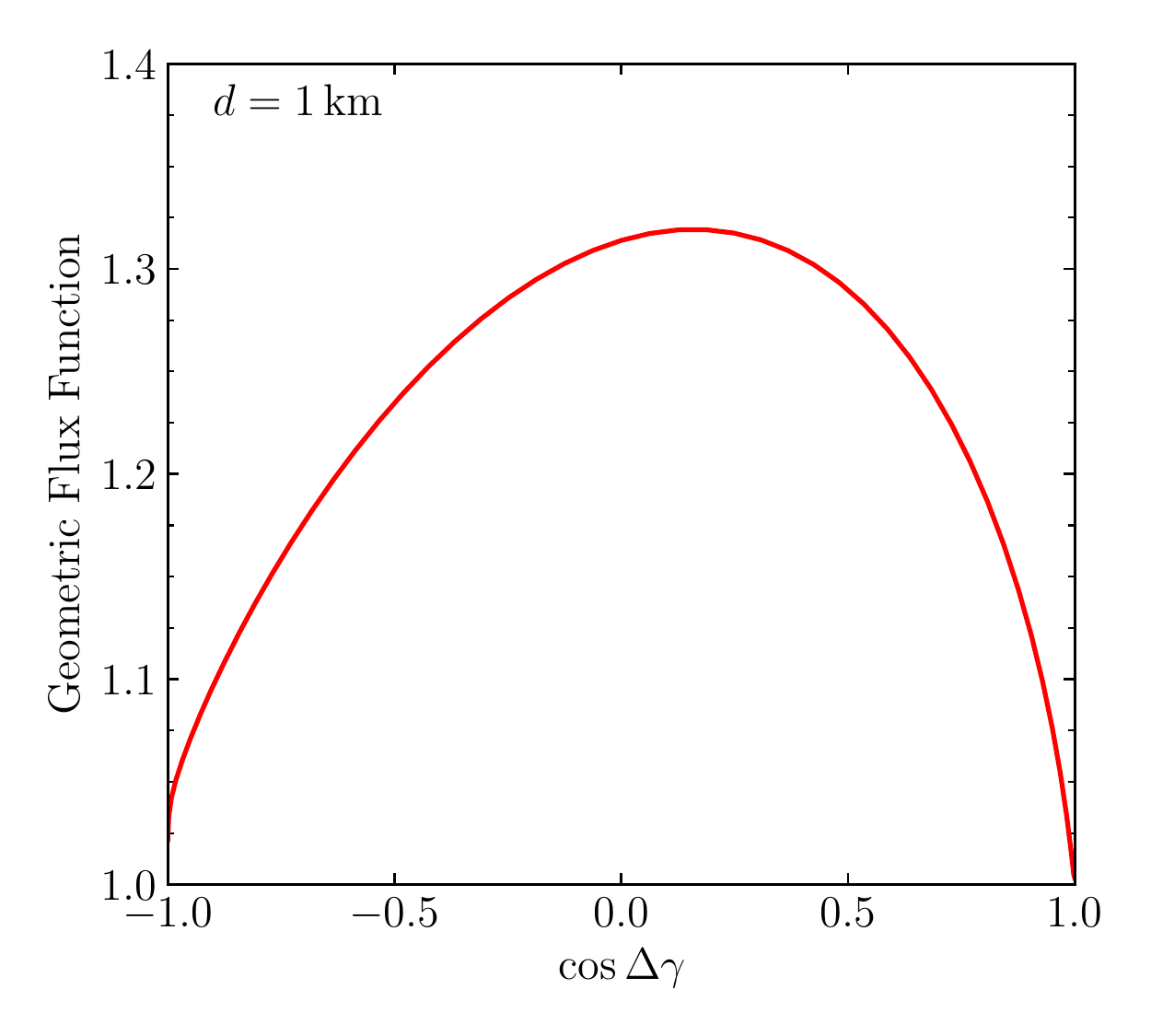}
    %\hspace{0.1cm}
  \end{center}
  \vspace{-0.2cm}
  \caption{(Left)~Geometry for our calculation of the neutrino flux in a detector at position $z = r_M-d$ in the Moon from cosmic rays hitting a surface patch at polar angle $\theta$. Here, $\hat{r}$ is the radial unit vector, $\hat{f}$ is the unit vector antiparallel to the incident CR direction, $\hat{D}$ is the unit vector pointing out from the detector to the patch, $\alpha$ is the polar angle between $\hat{r}$ and $\hat{D}$, and $\gamma$ is the angle between $\hat{f}$ and $\hat{D}$. Neutrinos are assumed to be emitted uniformly within a cone of opening angle $\Delta\gamma$ relative to the incident cosmic ray direction. (Right)~Variation in the geometric flux function at a depth of $d=r_M-z = 1\,\text{km}$ for different values of the neutrino emission cone angle $\Delta\gamma$.
  } 
    \label{fig:fbeam}
\end{figure}

In the analysis above we computed the cosmic ray~(CR) induced neutrino flux at a depth $d$ below the surface of the Moon assuming an average isotropic emission of neutrinos from each cosmic ray collision. Here, we generalize this calculation to the case where the neutrinos created are correlated with the incident CR direction. For a given neutrino emission energy $E_\nu$, we assume an angular distribution $\mathcal{F}(c_\gamma)$, where $\gamma$ is the polar angle of the neutrino with respect to the CR direction, $c_\gamma = \cos\gamma$, and we normalize such that $1=\int_{-1}^1\!dc_\gamma\,\mathcal{F}(c_\gamma)$.

To obtain the neutrino flux at a depth $d = r_M-z$, we consider a patch of lunar surface with polar angle $\theta$ as shown in the left panel of Fig.~\ref{fig:fbeam}. For this patch, we take $\hat{r}$ as the usual radial unit vector, $\hat{f}$ as the unit vector antiparallel to the incident CR direction, and $\hat{D}$ as the unit vector along the direction from the detector to the patch on the surface. Each patch has area $dA = 2\pi\,r_M^2dc_\theta$, while neutrinos emitted from the patch are diluted by the factor $1/2\pi D^2$ by the time they reach the detector. Summing over all incident CR directions and then over the lunar surface area, and treating the incident CR flux as isotropic with $d\Phi_{\rm CR}/d\Omega_{\rm CR} = \Phi_{\rm CR}/4\pi$, we have
\beq
\frac{d\Phi_\nu}{dE_\nu} = \frac{dN_\nu}{dE_\nu}\,{\Phi_{\rm CR}}\times\frac{1}{4\pi}\,\int_{-1}^1\!dc_\theta\,\frac{r^2}{D^2}
\int_{c_{\Delta\gamma}}^1\!\!dc_\gamma
\int_0^{2\pi}\!d\beta\;
(\hat{r}\cdot\hat{f})\,\Theta(\hat{r}\cdot\hat{f})\;
\mathcal{F}(c_\gamma)
\ .
\label{eq:fluxbeam}
\eeq
Here, $\gamma$ and $\beta$ are the polar and azimuthal coordinates of $\hat{f}$ relative to $\hat{D}$, the factor of $(\hat{r}\cdot\hat{f})$ projects onto the component normal to the surface, and the step function enforces the constraint that only downward-going CRs contribute. Limiting the integration to $c_\gamma \geq c_{\Delta\gamma}$ ensures that the detector lies within the emission cone of the incident CR. We also have
\beq
\hat{r}\cdot\hat{f} = c_\alpha c_\gamma + s_\alpha s_\gamma c_\beta \ ,
\eeq
where $\alpha$ is the angle between $\hat{r}$ and $\hat{D}$ as well as
\beq
\frac{r_M}{D} = \frac{1}{\sqrt{1+a^2-2ac_\theta}} \ ,\qquad 
c_\alpha = \frac{1-ac_\theta}{\sqrt{1+a^2-2ac_\theta}} \ ,
\eeq
with $a = z/r_M < 1$. From Eq.~\eqref{eq:fluxbeam} we see that the Moon CR neutrino flux can be written in terms of the universal production factor $\Phi_{\rm CR}\,dN_\nu/dE_\nu$ times a geometric flux function that depends on the depth $d$. 

Let us now specialize to a concrete angular distribution to illustrate the effect of deviating from the isotropic emission approximation. We consider
\beq
\mathcal{F}(c_\gamma) = \frac{1}{(1-c_{\Delta \gamma})}\;\Theta(c_\gamma- c_{\Delta\gamma}) \ ,
\eeq
corresponding to the uniform emission of neutrinos over a cone with polar angle $\Delta\gamma$. More general angular distributions can be built up by superposing such emission cones. In the right panel of Fig.~\ref{fig:fbeam} we show the variation in the geometric flux function at $d=r_M-z = 1\,\text{km}$ as $c_{\Delta\gamma}$ is varied over all possible values. For $c_{\Delta\gamma}\to -1$ we recover the result of Eq.\eqref{eq:mcrflux}, while for $c_{\Delta \gamma} \to 1$ the geometric function goes to unity as expected. From this figure we see that the effect of deviating from the isotropic emission approximation used in our main analysis only changes the neutrino flux at $d=1\,\text{km}$ by less than about $30\%$. %We note that this is something of a numerical accident, and that the variation at greater depths with $d\sim r_M$ is much larger. 
Note that this is only one of several uncertainties in our calculation of the CR neutrino flux on the Moon; like the others, it can be systematically improved through further simulations and measurements.

\subsection{Impacts of Flux Uncertainties and Detector Parameters}

The dark matter sensitivities shown in Fig.~\ref{fig:nufloor1} involve a number of well-motivated estimates about uncertainties in the neutrino fluxes and properties of the detectors. We examine here two examples of the impact of deviating from these estimates on the DM sensitivities, and we show that the results in the main text are robust under the variations.

\begin{figure}[ttt]
  \begin{center}
    \includegraphics[width = 0.47\textwidth]{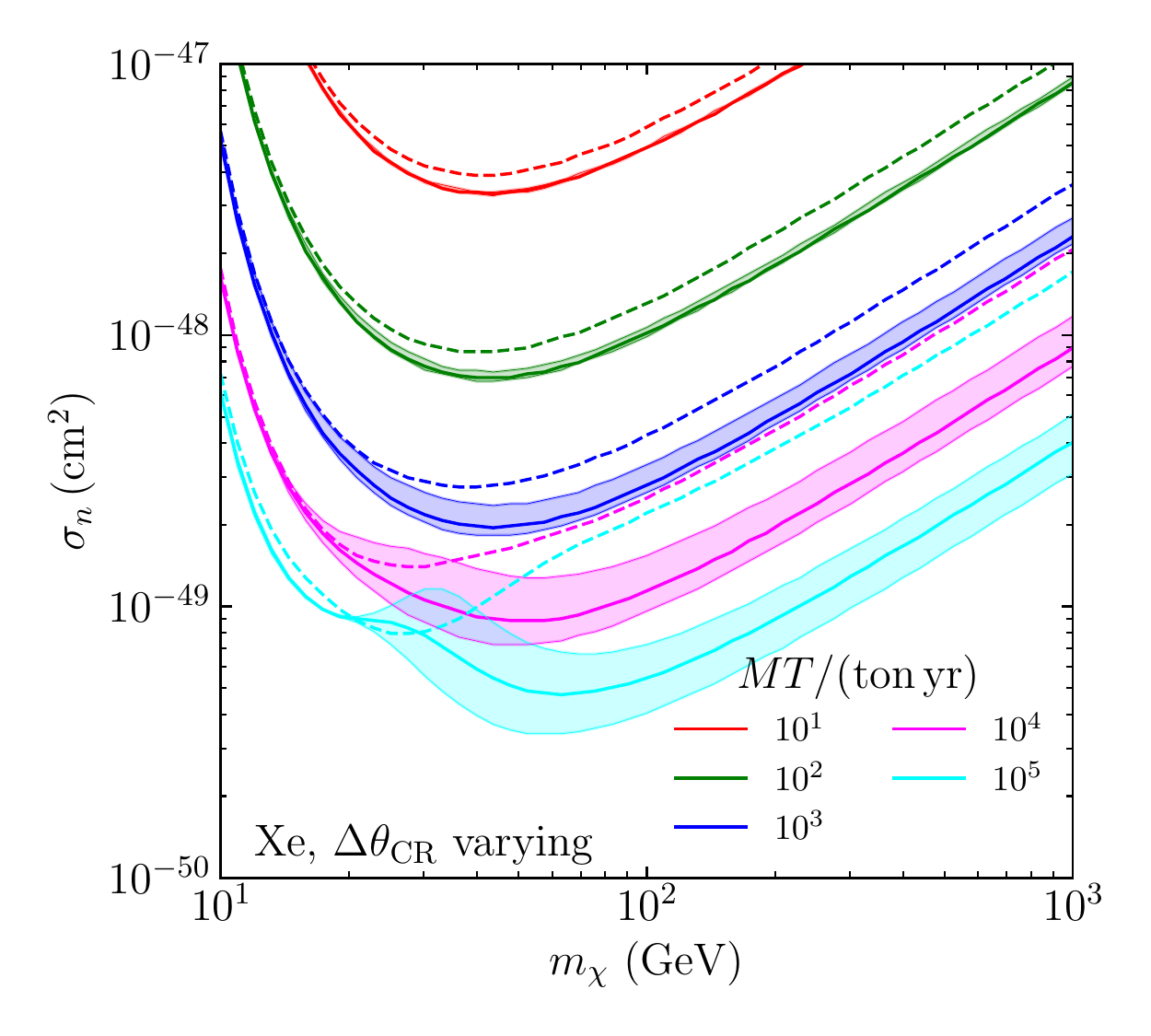}
    \includegraphics[width = 0.47\textwidth]{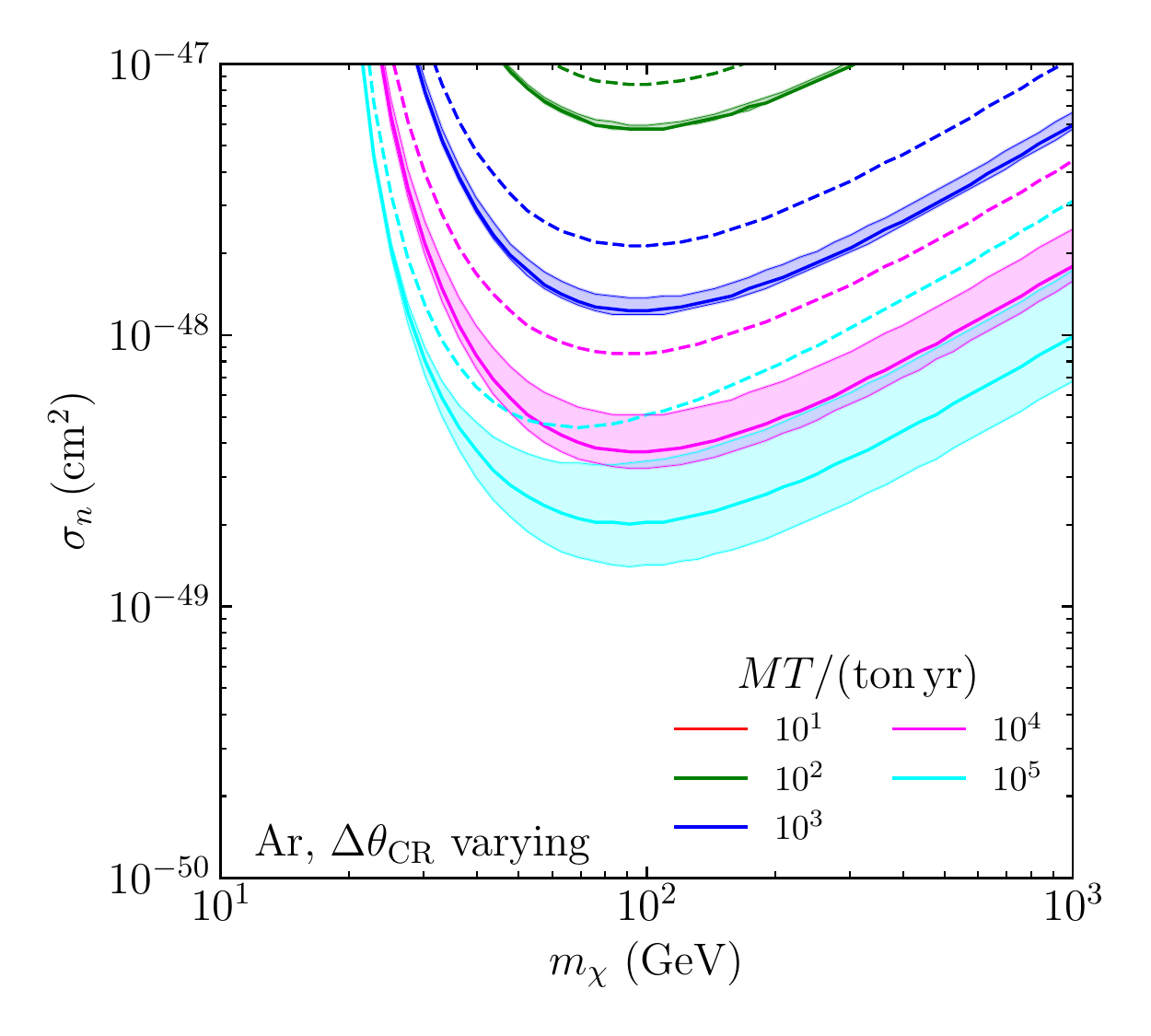}
    %\hspace{0.1cm}
  \end{center}
  \vspace{-0.5cm}
  \caption{Impact on DM sensitivity in xenon~(left) and argon~(right) of varying the fractional uncertainty in the Moon CR neutrino flux between $\Delta\theta_{CR} = 10$--$40\%$. The sensitivities are given in terms of the spin-independent dark matter per-nucleon cross section $\sigma_n$ as a function of DM mass $m_\chi$. The shaded regions bounded by solid lines indicate the sensitivities for a detector on the Moon as $\Delta\theta_{\rm CR}$ is varied over the range considered, while the dashed lines indicate the sensitivity on the Earth. The various colours from top to bottom denote total exposures of $MT/(\text{ton\,yr}) = 10^2,\,10^{3},\,10^{4},\,10^{5}$.
  }
  \label{fig:nufloor2}
\end{figure}

As a first example, recall that we assumed a 20\% uncertainty in the CR neutrino flux on the Moon. This is the same level of fractional uncertainty as expected for atmospheric fluxes on the Earth~\cite{Battistoni:2005pd,Barr:2006it,Honda:2011nf,Honda:2015fha}, and is realistically achievable by extending the initial calculations of this work. Even so, we consider the effect of different values of the fractional Moon CR flux uncertainty between $10$--$40\%$. This is shown in Fig.~\ref{fig:nufloor2}, where we plot the sensitivity to the DM per-nucleon cross section $\sigma_n$ as a function of mass $m_\chi$ for a set of total exposures $MT/(\text{ton yr})$ for representative xenon~(left) and argon~(right) detectors. The dashed lines show the sensitivities on the Earth and the solid lines the sensitivities on the Moon. The shaded bands for the Moon correspond to varying the Moon CR neutrino flux uncertainty between $10$--$40\%$. In general, the larger the uncertainty in the CR flux, the lower the sensitivity  to DM, although there is still a significant improvement in sensitivity on the Moon relative to the Earth over most of the ranges we consider.

\begin{figure}[t!]
  \begin{center}
    \includegraphics[width = 0.47\textwidth]{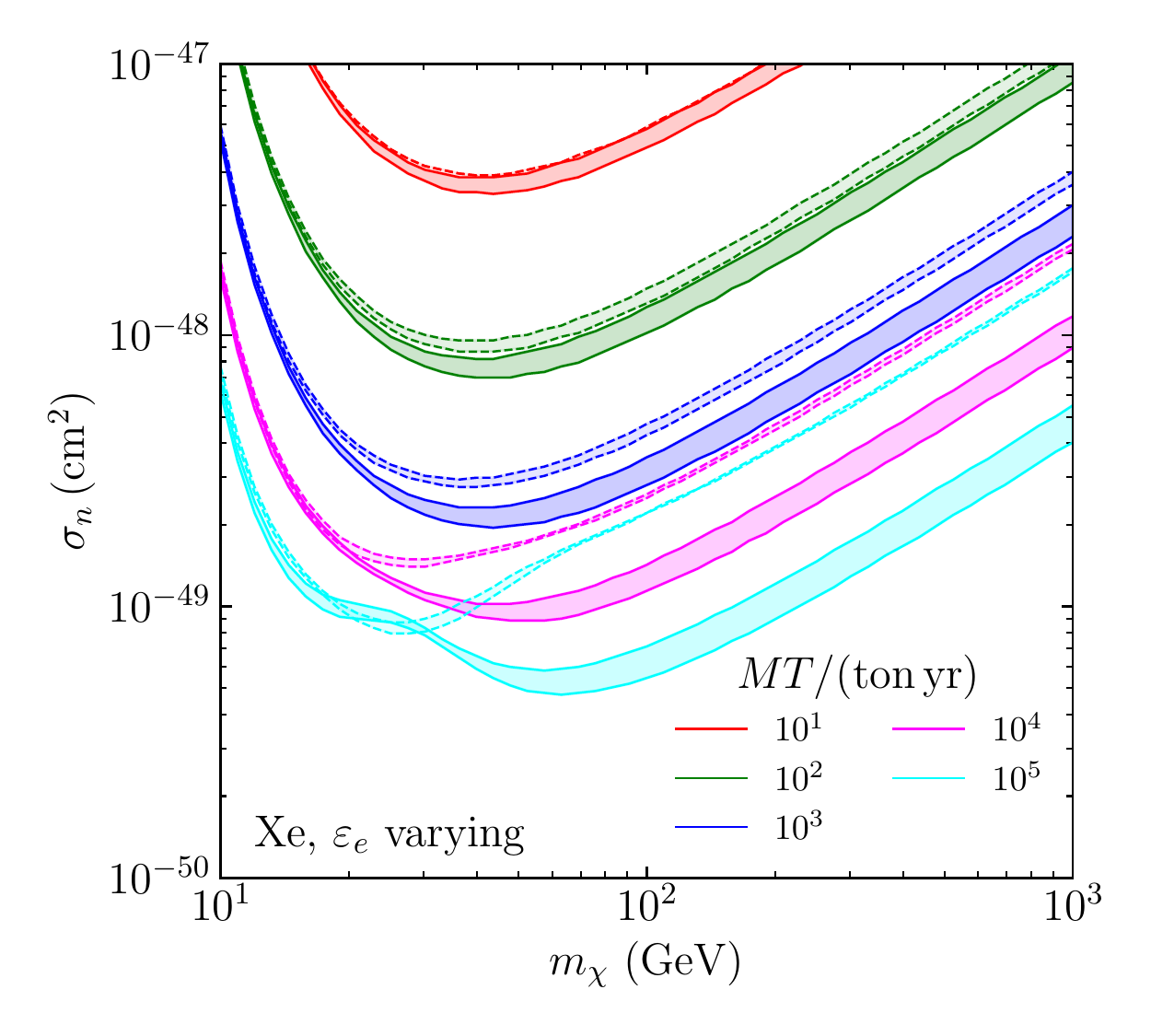}
  \end{center}
  \vspace{-0.5cm}
  \caption{Impact on DM sensitivity in xenon of varying the electron rejection factor between $\varepsilon_e = 2\times 10^{-4}$--$1\times 10^{-3}$. The sensitivities are given in terms of the spin-independent dark matter per-nucleon cross section $\sigma_n$ as a function of DM mass $m_\chi$. The shaded regions bounded by solid lines indicate the sensitivities for a detector on the Moon as $\varepsilon_e$ is varied over the range considered, while the shaded regions bounded by dashed lines show the corresponding sensitivity on the Earth. The various colours from top to bottom denote total exposures of $MT/(\text{ton\,yr}) = 10^2,\,10^{3},\,10^{4},\,10^{5}$.
  }
  \label{fig:nufloor3}
\end{figure}

The second example we consider is the impact of variations in the electron recoil rejection factor~(ERF) in xenon detectors relative to the value $\varepsilon_e=2\times 10^{-4}$ used in the analysis above. Distinguishing between electron and nuclear recoils in two-phase xenon detectors is achieved by comparing scintillation and ionization within each event. The target electron recoil rejection factor for DARWIN is 
$\varepsilon_e=2\times 10^{-4}$~\cite{DARWIN:2016hyl}, which is challenging but potentially achievable. In Fig.~\ref{fig:nufloor3} we show the DM discovery sensitivity for electron rejection values between $\varepsilon_e = 2\times 10^{-4}$--$1\times 10^{-3}$, corresponding to the shaded bands around the solid lines. As the electron rejection factor gets larger, more neutrino-electron events contribute to the DM background, and we see reduction in DM sensitivity. Even with this reduction, the sensitivities on the Moon remain considerably better than on the Earth at larger DM mass.
Note that no such issue is expected for argon detectors that use pulse-shape discrimination to reject electron recoils with $\varepsilon_e \lesssim 10^{-8}$~\cite{Boulay:2006mb,DEAP:2009hyz,Adhikari:2020zyy,DEAP:2021axq}.

\subsection{Dark Matter and Neutrino Scattering}

In many realizations of dark matter~(DM), the dominant interaction between dark matter and nuclei is spin-independent~(SI) and mediated by massive mediator particles. In this case, cross section per unit nuclear recoil energy $E_R$ for a DM particle $\chi$ to scatter off a nucleus $N=(A,Z)$ is
\beq
\frac{d\sigma_{\chi N}}{dE_R} = \frac{m_N}{2\mu_{N}^2v^2}\,\bar{\sigma}_{N}|F_N(E_R)|^2 \ ,
\label{eq:sigmadm}
\eeq
where $\mu_{N} = m_\chi m_N/(m_N+m_\chi)$, $v$ is the DM velocity in the lab frame, $m_N$ is the nuclear mass, $\bar{\sigma}_N$ depends on the target but is independent of $v$ and $E_R$, and $F_N$ is a nuclear form factor. For this last quantity, we use the Helm form~\cite{Engel:1992bf}: 
\beq
F_N(E_R) = \frac{3\left[\sin(qr_N)-(qr_N)\cos(qr_N)\right]}{(qr_N)^3}\,e^{-(qs)^2/2} \ ,
\eeq
where $q=\sqrt{2m_NE_R}$, $s=0.9\,\text{fm}$, and $r_N=\sqrt{c^2+(7\pi/3)\bar{a}^2-5s^2}$ with $c=(1.23a^{1/3}-0.6)\,\text{fm}$ and $\bar{a} = 0.52\,\text{fm}$.
To allow a comparison of direct DM searches using different nuclear targets, it is standard practice to define an effective per-nucleon DM cross section for SI scattering by
\beq
\sigma_n \equiv \frac{1}{A^2}\frac{\mu_n^2}{\mu_N^2}\bar{\sigma_N} \ , 
\eeq
where $\mu_n$ is the DM-nucleon reduced mass and $\mu_N$ is the DM-nucleus reduced mass. We express the sensitivity of prospective DM detectors in terms of $\sigma_n$ in the text.

The cross section for low-energy neutrino scattering on a nucleus $N=(A,Z)$ is given by~\cite{Freedman:1977xn}
\beq
\frac{d\sigma_{\nu N}}{dE_R} = \frac{G_F^2Q_W^2}{4\pi}m_N\left(1-\frac{m_NE_R}{E_\nu^2}\right)|F_N(E_R)|^2 \ ,
\label{eq:sigmanun}
\eeq
where $E_\nu$ is the neutrino energy, $E_R$ is the nuclear recoil energy, $G_F$ is the Fermi constant, $Q_W = (A-Z)-Z(1-s_W^2)$ with $s_W^2\simeq 0.23$, and $F_N$ is the same form factor as for SI DM scattering to an excellent approximation~\cite{Engel:1992bf,Vogel:1989iv}.  Note that kinematics implies that this cross section is non-zero only for $E_\nu > \sqrt{m_NE_R/2}$.

Neutrinos can also scatter on electrons in the target material leading to energy deposition that is sometimes mistaken for the nuclear recoils being searched for. For a free electron, the cross section for scattering by a neutrino of flavor $a$ per unit energy transfer $T$ is~\cite{Sarantakos:1982bp,Vogel:1989iv,Marciano:2003eq,Formaggio:2013kya}
\beq
\frac{d\sigma_{a}^{(e)}}{dT} = \frac{2G_F^2}{\pi}\,m_e\left[Q_+^2+Q_+^2\left(1-\frac{T}{E_\nu}\right)^2-Q_-Q_+\frac{m_eT}{E_\nu^2}\right] \ ,
\label{eq:sigmanue}
\eeq
with $Q_+ = s_W^2$ and $Q_+ = \delta_{ae}-1/2+s_W^2$. Kinematics requires $E_\nu > [T+\sqrt{T(T+m_e)}]/2$ for this expression to be non-zero. 

This free-electron cross section is a helpful reference for the quantity that we need; the cross section for neutrino scattering on the $Z$ atomic electrons in a target with nucleus $N=(A,Z)$. The dark matter detectors we investigate are sensitive to neutrino-electron scattering with $T\sim 1$--$20\,\kev$, and the dominant neutrino sources for this range of energy transfers are $pp$ and $^7\text{Be}$ neutrinos with $E_\nu \sim 100$--$1000\,\kev$~\cite{Billard:2013qya,Chen:2016eab}. Since these energies typically are large compared to typical electron binding energies and the energy transfer, atomic effects are expected to only give a moderate correction to simply rescaling the free electron cross section of Eq.~\eqref{eq:sigmanue} by the number of electrons $Z$ in the atom~\cite{Voloshin:2010vm}. For argon, we apply a slightly improved version of this approximation and take~\cite{Kouzakov:2011ka,Kouzakov:2011vx,Kouzakov:2014lka}
\beq
\frac{d\sigma^{(\rm Ar)}_{a}}{dT} = \sum_n\Theta(T-E_n)\,\frac{d\sigma_{a}^{(e)}}{dT} \ ,
\eeq
where the sum runs over all energy levels $E_n$ in argon. In xenon, the binding energies are larger and atomic effects are more important, so for this element we apply the detailed atomic calculation of Ref.~\cite{Chen:2016eab}.

\end{document}